\documentclass{WileyMSP-template}
\usepackage{cite}
\usepackage{amssymb}
\usepackage{amsmath}
\usepackage{graphicx}
\usepackage{txfonts}
\usepackage{color}
\usepackage{amsfonts}
\usepackage[colorlinks,urlcolor=blue,linkcolor=blue,citecolor=blue]{hyperref}

\begin{document}

\pagestyle{fancy}
\rhead{\includegraphics[width=2.5cm]{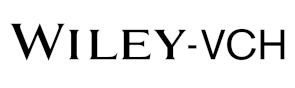}}

\title{Two-mode correlated  multiphoton bundle emission}

\maketitle


\author{Yi Wang}
\author{Fen Zou}
\author{Jie-Qiao Liao*}\\
\begin{affiliations}
Y. Wang, J.-Q. Liao\\
Key Laboratory of Low-Dimensional Quantum Structures and Quantum Control of
Ministry of Education, Key Laboratory for Matter Microstructure and Function of Hunan Province, Department of Physics and Synergetic Innovation Center for Quantum Effects and Applications, Hunan Normal University, Changsha 410081, China\\
J.-Q. Liao\\
Institute of Interdisciplinary Studies, Hunan Normal University, Changsha, 410081, China\\
Email Address:jqliao@hunnu.edu.cn\\
F. Zou\\
Beijing Computational Science Research Center, Beijing 100193, China
\end{affiliations}



\begin{abstract}
The preparation of correlated multiphoton sources is an important research topic in quantum optics and quantum information science. Here, two-mode correlated multiphoton bundle emission in a nondegenerate multiphoton Jaynes-Cummings model, which is comprised of a two-level system coupled with two cavity modes is studied. The two-level system is driven by a near-resonant strong laser such that the Mollow regime dominates the physical processes in this system. Under certain resonance conditions, a perfect super-Rabi oscillation between the zero-photon state $|0\rangle_{a}|0\rangle_{b}$ and the ($n+m$)-photon state $|n\rangle_{a}|m\rangle_{b}$ of the two cavity modes can take place. Induced by the photon decay, the two-mode correlated multiphoton bundle emission occurs in this system. More importantly, the results show that there is an antibunching effect between the strongly-correlated photon bundles, so that the system behaves as an antibunched ($n+m$)-photon source. The work opens up a route towards achieving two-mode correlated multiphoton source device, which has potential applications in modern quantum technology.
\end{abstract}


\section{Introduction}
Multiphoton sources\textsuperscript{\cite{Dell20o6,Pan2012}} are important resources in quantum information science, and have wide applications in many frontier research fields, including quantum metrology,\textsuperscript{\cite{Giovannetti2004,Giovannetti2006}} quantum lithography,\textsuperscript{\cite{PhysRevLett2000,Two-photon2001}} quantum spectroscopy,\textsuperscript{\cite{López2015,Dorfman2016}} quantum imaging,\textsuperscript{\cite{Moreau2019}} and quantum communication.\textsuperscript{\cite{Kimble2008Jun,Gisin2006,Gisin2002}} Currently, great advances have been made in the preparation of multiphoton sources, and a variety of schemes for creating multiphoton sources have been proposed in various physical platforms, such as photonic waveguides,\textsuperscript{\cite{alez-Tudela2017}} Rydberg atomic systems,\textsuperscript{\cite{Bienias2014,Maghrebi2015}} cavity quantum electrodynamics (QED) systems,\textsuperscript{\cite{Law1996,Huang2014}} multi-level atomic systems,\textsuperscript{\cite{Chang2016,Photonics2014,Hargart2016,Koshino2013,Dousse2010,Iwamoto2011,Zueco2016}} Kerr nonlinear cavities,\textsuperscript{\cite{Liao2010}} and cavity optomechanical systems.\textsuperscript{\cite{Liao2013,Qin2019}} Multiphoton bundle emission\textsuperscript{\cite{Laussy2014,Bin2021,Ma2021,Cosacchi2021,Ren2022,Jiang2022,liu2022,Zou2023,Gou2022,SanchezMunoz2018,Ma2022}} is considered a reliable physical mechanism for the creation of multiphoton sources. Recently, the generation of $N$-photon ($N$-phonon) bundle emission has been studied in cavity-QED systems,\textsuperscript{\cite{Laussy2014,Ren2022,Jiang2022,Gou2022,Cosacchi2021,liu2022,Bin2021,SanchezMunoz2018,Bin2020,Deng2021}} circuit-QED systems,\textsuperscript{\cite{Ma2021,Zou2023,Ma2022}} and cavity optomechanical systems.\textsuperscript{\cite{Zou2022}} Notably, the $N$-photon bundle emission has been successfully accomplished in a dc-biased superconducting circuit.\textsuperscript{\cite{Menard2022}} The characteristics of $N$-photon bundle emission can be controlled by tuning the system parameters, enabling the realization of both $N$-photon lasers and $N$-photon guns.\textsuperscript{\cite{Laussy2014}} In addition, multiphoton sources can be  achieved through the physical mechanism of the multiphoton blockade effect.\textsuperscript{\cite{Ridolfo2012,Huang2018,Miranowicz2013,Zou2020,Ren2021}}

The nondegenerate multiphoton Jaynes-Cummings (JC) model,\textsuperscript{\cite{Ashraf1992,Ashraf1994}} as one of the important physical models in quantum optics, plays a crucial role in the study of quantum entanglement and quantum information processing. In this paper, we study the two-mode correlated multiphoton bundle emission in a nondegenerate multiphoton JC model consisting of a two-level system (TLS) and two cavity modes. Here the TLS is coupled to the two cavity modes via the multiphoton JC interaction and the TLS is driven by a near-resonant strong laser. By analyzing the eigenenergy spectrum of the system in the Mollow regime,\textsuperscript{\cite{Mollow1969,Kimble1976,Ulhaq2012,Gonzalez-Tudela2013,Carreño2017}} we find that, under suitable resonance conditions, a transition from the zero-photon state $|0\rangle_{a}|0\rangle_{b}|+\rangle$ to the ($n+m$)-photon state $\vert n\rangle_{a}\vert m\rangle_{b}\vert-\rangle$ can be achieved, where $\vert n\rangle_{a}$ and $\vert m\rangle_{b}$ ($n, m$ are natural numbers) are the number states of the two cavity modes and $|\pm\rangle$ are the eigenstates of the driven TLS in the absence of the JC interaction. To provide the evidence for achieving a perfect super-Rabi oscillation between the zero-photon state $\vert0\rangle_{a}\vert0\rangle_{b}\vert+\rangle$ and the ($n+m$)-photon state $\vert n\rangle_{a}\vert m\rangle_{b}\vert-\rangle$ ($n, m=1, 2$), we examine the populations of the two states. Based on such a super-Rabi oscillation, the two-mode correlated multiphoton bundle emission can be further achieved when opening a photon decay channel. To analyze the two-mode correlated multiphoton bundle emission and investigate the quantum statistics between the photon bundles, we calculate the joint two-mode photon number distribution, the equal-time high-order correlation function between the two cavity modes, the generalized time-delayed second-order correlation functions of the photon bundle, and the quantum trajectory of the state populations. The results show that two-mode multiphoton bundle emission can be achieved in the nondegenerate multiphoton JC model, and the anti-bunching effect among the strongly-correlated photon bundles can be observed.

The rest of this paper is organized as follows: In Section \ref{sectionII}, we introduce the physical model and present the Hamiltonian. In Section \ref{sectionIII}, we study the eigen-system of the Hamiltonian in the Mollow regime and discuss the super-Rabi oscillation between the zero-photon state $\vert0\rangle_{a}\vert0\rangle_{b}\vert+\rangle$ and the ($n+m$)-photon state $\vert n\rangle_{a}\vert m\rangle_{b}\vert-\rangle$ for $n, m=1, 2$ as examples. We also study the two-mode correlated multiphoton bundle emission by calculating the joint two-mode photon number distribution and correlation functions. In addition, we show a quantum Monte-Carlo simulation of the multiphoton bundle emission processes. In Section \ref{sectionIV}, we present some discussions on the experimental feasibility of the theoretical scheme and conclude this work. A detailed derivation of the effective oscillating frequency of the super-Rabi oscillation is presented in the Appendix.

\section{Model}\label{sectionII}
\begin{figure*}[t]
	\centering
	\includegraphics[width=0.79 \textwidth]{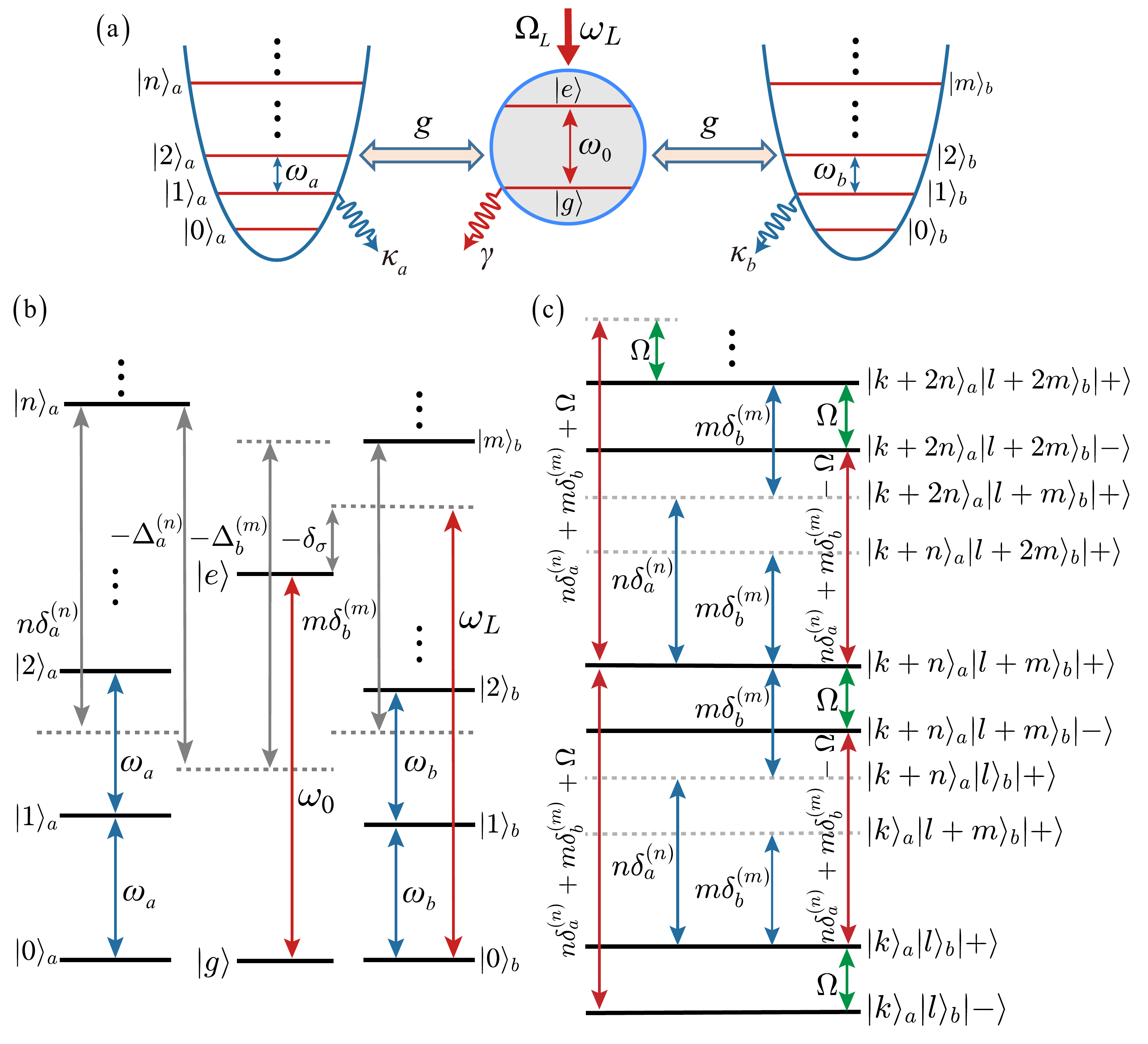}
	\caption{a) Schematic of the nondegenerate multiphoton JC model composed of a TLS coupled to two optical modes via the nondegenerate multiphoton JC process. b) Energy-level diagram of the bare states of the TLS and modes $a$ and $b$ in the Schr\"{o}dinger picture. c) Energy levels of the system in the Mollow regime. Here, the energy levels are expressed in the eigen-representation of the Hamiltonian $\hat{H}_{0}^{\prime}$ in Equation (\ref{H_{0}}), and the transitions occur among the ($k+l$)-, $(k+l+n+m)$-, and $(k+l+2n+2m)$-photon domains.}\label{Fig1}
\end{figure*}

We consider a two-mode nondegenerate multiphoton JC model consisting of a TLS coupled to two cavity modes, as shown in Figure {\ref{Fig1}}a. This model is described by the Hamiltonian ($\hbar=1$)
\begin{equation}
	\hat{H}_{\textrm{mpJC}}=\frac{\omega _{0}}{2}\hat{\sigma }_{z}+\omega_{a}\hat{a}^{\dagger}\hat{a}+\omega_{b}\hat{b}^{\dagger}\hat{b}+g(\hat{a}^{n}\hat{b}^{m}\hat{\sigma } _{+}+\hat{\sigma }_{-}\hat{a}^{\dagger n}\hat{b}^{\dagger m}), \label{1}
\end{equation}   
where $\hat{a}~(\hat{a}^{\dagger})$ and $\hat{b}~(\hat{b}^{\dagger})$ are the annihilation (creation) operator of modes $a$ and $b$ with resonance frequencies $\omega_{a}$ and $\omega_{b}$, respectively. The $\hat{\sigma}_{+}=\hat{\sigma}_{-}^{\dagger}=|e\rangle\langle g|=(\hat{\sigma}_{x}+i\hat{\sigma}_{y})/2$ is the raising operator of the TLS with the transition frequency $\omega _{0}$ between the excited state $|e\rangle $ and the ground state $|g\rangle $, where $\hat{\sigma }_{x}=|e\rangle\langle g|+|g\rangle \langle e|, \hat{\sigma }_{y}=-i(|e\rangle\langle g|-|g\rangle \langle e|),$ and $\hat{\sigma }_{z}=|e\rangle\langle e|-|g\rangle \langle g|$ are the Pauli operators of the TLS. The parameter $g$ is the coupling strength of the multiphoton JC interaction, and $n~(m)$ is the photon number of mode $a$ (mode $b$) associated with each transition process of the TLS. Here, the variables $n$ and $m$ take natural numbers and satisfy the relation $n+m>0$.

We assume that the TLS is driven by a strong laser with frequency $\omega_{L}$ and amplitude $\Omega_{L}$, then the total Hamiltonian of the system reads
\begin{align}
	\hat{H}_{\textrm{sys}}=&\hat{H}_{\textrm{mpJC}}+\Omega_{L}(\hat{\sigma}_{+}e^{-i\omega_{L}t}+\hat{\sigma}_{-}e^{i\omega_{L}t}).\label{2}
\end{align}
In a rotating frame defined by the unitary operator $\exp\{-i\omega_{L}t [\hat{\sigma}_{z}/2+\hat{a}^{\dag}\hat{a}/(2n)+\hat{b}^{\dag }\hat{b}/(2m)]\}$, the Hamiltonian $\hat{H}_{\textrm{sys}}$ becomes
\begin{equation}
	\begin{split}
		\hat{H}_{I}=&\frac{\delta_{\sigma}}{2}\hat{\sigma}_{z}+\delta_{a}^{(n)}\hat{a}^{\dagger}\hat{a}+\delta_{b}^{(m)}\hat{b}^{\dagger}\hat{b}+g(\hat{a}^{n}\hat{b}^{m}\hat{\sigma}_{+}+\hat{\sigma}_{-}\hat{a}^{\dagger n}\hat{b}^{\dagger m})+\Omega_{L}\hat{\sigma}_{x},\label{HI}
	\end{split}	
\end{equation}
where $\delta_{a}^{(n)}=\omega_{a}-\omega_{L}/(2n)$ is the $n$-photon transition detuning in mode $a$, $\delta_{b}^{(m)}=\omega_{b}-\omega_{L}/(2m)$  is the $m$-photon transition detuning in mode $b$, and $\delta_{\sigma}=\omega_{0}-\omega_{L}$ is the atomic driving detuning, as shown in Figure \ref{Fig1}b.
\section{Two-mode correlated multiphoton bundle emission in the Mollow regime}\label{sectionIII}
In order to achieve two-mode correlated multiphoton bundle emission, it is typically necessary to generate the corresponding two-mode multiphoton states in advance. In this nondegenerate multiphoton JC model, the two-mode multiphoton state (i.e., $|n\rangle_{a}|m\rangle_{b}$) can be generated based on the physical mechanism of super-Rabi oscillation between the two states $|0\rangle_{a}|0\rangle_{b}$ and $|n\rangle_{a}|m\rangle_{b}$. In this section, we discuss the realization of the super-Rabi oscillation for $n, m=1, 2$ in this system. We also study the two-mode correlated multiphoton bundle emission by calculating the joint two-mode photon number distribution, the equal-time high-order correlation functions of the two different cavity modes, the generalized time-delayed second-order correlation function of multiphoton bundle, and the quantum Monte-Carlo simulation of multiphoton bundle emission.
\subsection{Super-Rabi oscillation}\label{subsectionA}
In previous works on bundle emission, it has been shown that, for the single-mode ($n$-photon) JC model, the super-Rabi oscillation can be realized in both the Mollow regime and the  ($n$-photon) JC-coupling regime\textsuperscript{\cite{Laussy2014,Jiang2022}}. In this paper, we focus on the Mollow regime of the nondegenerate multiphoton JC model, where the driving amplitude $\Omega_{L}$ is much stronger than the coupling strength $g$ of the multiphoton JC interaction, that is, $\Omega_{L}\gg g$. In this case, we treat the multiphoton JC-coupling term as a perturbation. In the absence of this perturbation, the system Hamiltonian becomes
\begin{equation}
	\hat{H}_{0}^{\prime}=\frac{\delta_{\sigma}}{2}\hat{\sigma}_{z}+\delta_{a}^{(n)}\hat{a}^{\dagger}\hat{a}+\delta_{b}^{(m)}\hat{b}^{\dagger}\hat{b}+\Omega_{L}\hat{\sigma}_{x}.\label{H_{0}}
\end{equation}
The eigen-system of the Hamiltonian $\hat{H}_{0}^{\prime}$ is given by 
\begin{equation}
	\hat{H}_{0}^{\prime}|n\rangle_{a}|m\rangle_{b}|\pm\rangle=(E_{\pm }+n\delta_{a}^{(n)}+m\delta_{a}^{(m)})|n\rangle_{a}|m\rangle_{b}|\pm\rangle,
\end{equation}
where $E_{\pm }=\pm\Omega/2$ and $|\pm\rangle=c_{\pm}|e\rangle \pm c_{\mp}|g\rangle$ are, respectively, the eigenvalues and eigenstates of the Hamiltonian $\hat{H}_{\sigma}=(\delta_{\sigma}/2)\hat{\sigma}_{z}+\Omega_{L}\hat{\sigma}_{x}$, i.e., $\hat{H}_{\sigma}|\pm\rangle=E_{\pm }|\pm\rangle$. Here, the superposition coefficients are given by $c_{\pm }=\sqrt{{2\Omega_{L}^{2}}/({\Omega^{2}\mp \Omega\delta_{\sigma})}}$, and the generalized Rabi frequency is introduced as $\Omega =\sqrt{\delta_{\sigma}^{2}+4\Omega_{L}^{2}}$. 

By using the completeness relation
\begin{align}
	\sum_{k,l=0}^{\infty}\sum_{s=\pm}|k\rangle_{a}|l\rangle_{b}\,_{a}\langle k|\,_{b}\langle l| \otimes |s\rangle \langle s|=\text{I},
\end{align}
we can expand the multiphoton JC coupling term [i.e., the $g$ term in Equation (\ref{HI})] with the eigenstates of the Hamiltonian $\hat{H}_{0}^{\prime}$ as
\begin{equation}	
	\begin{split}
		\hat{H}_{I}^{\prime}=&\sum_{k,l=0}^{\infty}\sum_{s,r=\pm}g\sqrt{\frac{(k+n)!}{k!}}\sqrt{\frac{(l+m)!}{l!}}A_{sr}\\
		&\times|k+n\rangle_{a}|l+m\rangle_{b} \,_{a}\langle k|\,_{b}\langle l|\otimes |s\rangle \langle r|+\text{H.c.},\label{HI1}
	\end{split}
\end{equation}
where $n$ and $m$ are the photon-number indexes defined in Equation (\ref{1}). In a rotating frame defined by the unitary operator $\hat{U}_{0}^{\prime}(t)=\exp(-i\hat{H}_{0}^{\prime}t)$, the Hamiltonian in
Equation (\ref{HI1}) becomes
\begin{eqnarray}
	\begin{split}
		\hat{V}_{I}=&\sum_{k,l=0}^{\infty}\sum_{s,r=\pm}B_{nmsr}A_{sr}|k+n\rangle_{a}|l+m\rangle_{b}\,_{a}\langle k|\,_{b}\langle l|\otimes |s\rangle \langle r|+\text{H.c.},\label{VI}
	\end{split}
\end{eqnarray}
where we introduce
\begin{subequations}
	\begin{align}
		B_{nmsr}&=g\sqrt{\frac{(k+n)!}{k!}}\sqrt{\frac{(l+m)!}{l!}}e^{i(E_{s}-E_{r}+n\delta_{a}^{(n)}+m\delta_{b}^{(m)})t },\\
		A_{++}&=\langle+|\sigma_{-}|+\rangle=c_{+}c_{-},\\ 
		A_{+-}&=\langle+|\sigma_{-}|-\rangle=c_{-}^{2},\\
		A_{-+}&=\langle-|\sigma_{-}|+\rangle=-c_{+}^{2} ,\\
		A_{--}&=\langle-|\sigma_{-}|-\rangle=-c_{+}c_{-}.
	\end{align}
\end{subequations}
Here, $B_{nmsr}$ and $A_{sr}$ are the transition amplitudes contributed by the two cavity modes and the TLS, respectively.
\begin{figure}
	\center
	\includegraphics[width=0.49 \textwidth]{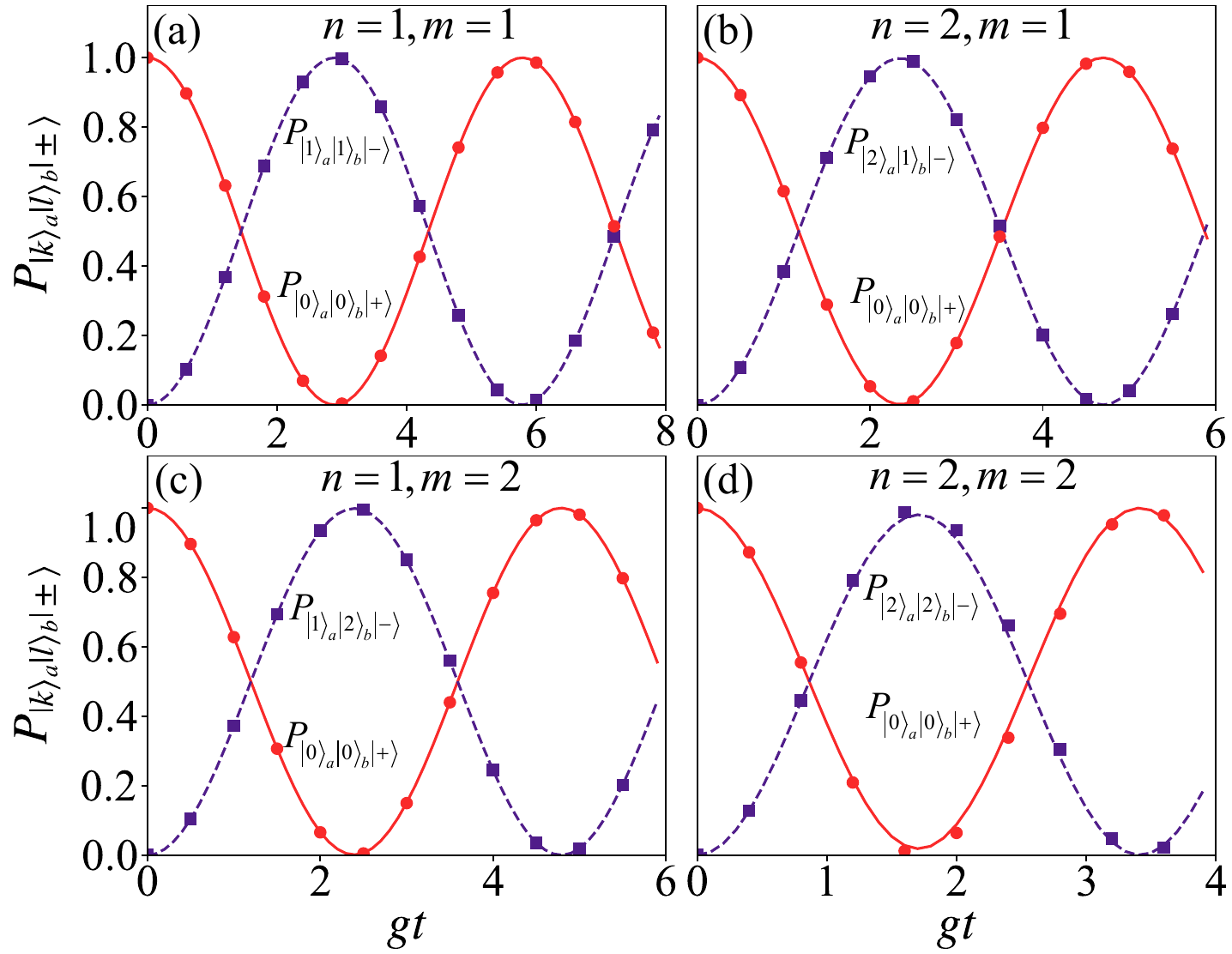}
	\caption{(Color online) The state populations $P_{|0\rangle_{a}|0\rangle_{b}|+\rangle}$ and $P_{|n\rangle_{a}|m\rangle_{b}|-\rangle}$ ($n, m=1, 2$) as functions of the scaled evolution time $gt$ for a) the nondegenerate two-photon JC model ($n=1, m=1$) at $\Delta_{a}^{(1)}/g=-55, \Delta_{b}^{(1)}/g=-70$, and $\Omega_{L}/g=90$; b) the nondegenerate three-photon JC model ($n=2, m=1$) at $\Delta_{a}^{(2)}/g=-115, \Delta_{b}^{(1)}/g=-70$, and $\Omega_{L}/g=100$; c) the nondegenerate three-photon JC model ($n=1, m=2$) at $\Delta_{a}^{(1)}/g=-55, \Delta_{b}^{(2)}/g=-145$, and $\Omega_{L}/g=100$; and d) the nondegenerate four-photon JC model ($n=2, m=2$) at $\Delta_{a}^{(2)}/g=-115, \Delta_{b}^{(2)}/g=-145$, and $\Omega_{L}/g=110$. The red solid and blue dashed curves correspond to the numerical results of the state populations, while the red dots and the blue rectangles correspond to the analytical results based on the effective Rabi frequencies $\Omega_{\text{eff}}^{(1,1)}$, $\Omega_{\text{eff}}^{(2,1)}$, $\Omega_{\text{eff}}^{(1,2)}$, and $\Omega_{\text{eff}}^{(2,2)}$ defined in Equation (\ref{Omega}). The detunings are taken as $\delta_{a}^{(n)}=-{[(\Delta_{b}^{(m)}-3\Delta_{a}^{(n)})(\Delta_{a}^{(n)}+\Delta_{b}^{(m)})-4\Omega_{L}^{2}}]/[4n(\Delta_{a}^{(n)}+\Delta_{b}^{(m)}))]$ and $\delta_{b}^{(m)}=-{[(\Delta_{a}^{(n)}-3\Delta_{b}^{(m)})(\Delta_{a}^{(n)}+\Delta_{b}^{(m)})-4\Omega_{L}^{2}}]/[4m(\Delta_{a}^{(n)}+\Delta_{b}^{(m)})]$.}\label{Fig2}
\end{figure}

It can be seen from Equation (\ref{VI}) that the nondegenerate multiphoton JC interaction will lead to a transition between the states $|k\rangle _{a}|l\rangle _{b}|r\rangle$ for $r=\pm$ and $|k+n\rangle _{a}|l+m\rangle _{b}|s\rangle$ for $s=\pm$, i.e., $|k\rangle _{a}|l\rangle_{b}|\pm \rangle \leftrightarrow |k+n\rangle _{a}|l+m\rangle _{b}|\pm\rangle$. This means that the two-mode multiphoton state $|n\rangle_{a}|m\rangle_{b}$ can be generated when the system is initially in the state $|0\rangle_{a}|0\rangle_{b}$. By analyzing the energy-levels  of the nondegenerate JC model in the Mollow regime, we know that the transition frequencies corresponding to the processes $|k\rangle_{a}|l\rangle_{b}|+\rangle \leftrightarrow |k+n\rangle _{a}|l+m\rangle _{b}|+\rangle$ and $|k\rangle_{a}|l\rangle_{b}|-\rangle \leftrightarrow |k+n\rangle_{a}|l+m\rangle_{b}|-\rangle$ are $n\delta_{a}^{(n)}+m\delta_{b}^{(m)}$. In this case, the high-order photon transitions will not be suppressed. Meanwhile, the transition frequency of the transition process $|k\rangle_{a}|l\rangle_{b}|+\rangle \leftrightarrow |k+n\rangle_{a}|l+m\rangle_{b}|-\rangle$ is $n\delta_{a}^{(n)}+m\delta_{b}^{(m)}-\Omega$. The corresponding high-order photon transition processes $|k+n\rangle_{a}|l+m\rangle_{b}|-\rangle \leftrightarrow |k+2n\rangle_{a}|l+2m\rangle_{b}|-\rangle$ and $|k+n\rangle_{a}|l+m\rangle_{b}|-\rangle\leftrightarrow |k+2n\rangle_{a}|l+2m\rangle_{b}|+\rangle$ will be suppressed by the detunings $\Omega$ and $2\Omega$, respectively. Similarly, for the transition process $|k\rangle_{a}|l\rangle_{b}|-\rangle \leftrightarrow |k+n\rangle_{a}|l+m\rangle_{b}|+\rangle$, the high-order photon transition processes $|k+n\rangle_{a}|l+m\rangle_{b}|+\rangle \leftrightarrow |k+2n\rangle_{a}|l+2m\rangle_{b}|+\rangle$ and $|k+n\rangle_{a}|l+m\rangle_{b}|+\rangle\leftrightarrow |k+2n\rangle_{a}|l+2m\rangle_{b}|-\rangle$ are also suppressed by the detunings $\Omega$ and $2\Omega$, respectively. When the detuning $\Omega$ is much larger than the corresponding coupling strengths, the upper detuned transitions can be eliminated. Therefore, for the transition processes $|k\rangle_{a}|l\rangle_{b}|+\rangle \leftrightarrow |k+n\rangle_{a}|l+m\rangle_{b}|-\rangle $ and $|k\rangle_{a}|l\rangle_{b}|-\rangle \leftrightarrow |k+n\rangle_{a}|l+m\rangle_{b}|+\rangle$, the Hilbert space of the system can be truncated into the two-dimension subspaces with the bases $\{|k\rangle_{a}|l\rangle_{b}|+\rangle, |k+n\rangle_{a}|l+m\rangle_{b}|-\rangle\}$ and $\{|k\rangle_{a}|l\rangle_{b}|-\rangle, |k+n\rangle_{a}|l+m\rangle_{b}|+\rangle\}$, respectively. This truncation provides the physical mechanism for the realization of the expected super-Rabi oscillation.

Based on the above energy-level analyses, we obtain the resonant transition conditions associated with the transition processes $|k\rangle_{a}|l\rangle_{b}|-\rangle\leftrightarrow |k+n\rangle_{a}|l+m\rangle_{b}|+\rangle$ and $|k\rangle_{a}|l\rangle_{b}|+\rangle\leftrightarrow |k+n\rangle_{a}|l+m\rangle_{b}|-\rangle$ as $n\delta_{a}^{(n)}+m\delta_{b}^{(m)}+\Omega=0$ and $n\delta_{a}^{(n)}+m\delta_{b}^{(m)}-\Omega=0$, respectively. In terms of the relations $\Delta_{a}^{(n)}=\omega_{0}/2-n\omega_{a}$, $\Delta_{b}^{(m)}=\omega_{0}/2-m\omega_{b}$, and $\delta_{\sigma}=\Delta_{a}^{(n)}+\Delta_{b}^{(m)}+n\delta_{a}^{(n)}+m\delta_{b}^{(m)}$, the detunings $\delta_{a}^{(n)}$ and $\delta_{b}^{(m)}$ can be obtained as
\begin{subequations}
	\begin{align}
		\delta_{a}^{(n)}&=\frac{(\Delta_{b}^{(m)}-3\Delta_{a}^{(n)})(\Delta_{a}^{(n)}+\Delta_{b}^{(m)}) -4\Omega_{L}^{2}}{4n(\Delta_{a}^{(n)}+\Delta_{b}^{(m)})},\\
		\delta_{b}^{(m)}&=\frac{(\Delta_{a}^{(n)}-3\Delta_{b}^{(m)})(\Delta_{a}^{(n)}+\Delta_{b}^{(m)})-4\Omega_{L}^{2}}{4m(\Delta_{a}^{(n)}+\Delta_{b}^{(m)})},
	\end{align}
\end{subequations}
where $\Omega_{L}$ is the driving amplitude. Notably, the two-mode nondegenerate multiphoton JC model is reduced to the single-mode multiphoton JC model when either $m=0$ or $n=0$. Furthermore, when $m=n$ and the resonance frequencies ($\omega_a$ and $\omega_b$) of modes $a$ and $b$ are equal, the detunings $\Delta_a^{(n)}$ and $\Delta_b^{(m)}$ are identical and the transition conditions are equivalent to those in the single-mode multiphoton JC model.\textsuperscript{\cite{Jiang2022}} In addition, we also obtain the resonant transition conditions associated with the higher-order transitions $|0\rangle_{a}|0\rangle_{b}|+\rangle \leftrightarrow
|\mu n\rangle_{a}|\mu m\rangle_{b} |-\rangle$ and $|0\rangle_{a}|0\rangle_{b}|-\rangle \leftrightarrow |\mu n\rangle_{a}|\mu m\rangle_{b}|+\rangle$ ($\mu\geq2$, $\mu$ is integer) as $\mu n\delta_{a}^{(n)}+\mu m\delta_{b}^{(m)}-\Omega=0$ and $\mu n\delta_{a}^{(n)}+\mu m\delta_{b}^{(m)}+\Omega=0$, respectively. In these cases, the detunings are determined by
\begin{eqnarray}
	n\delta_{a}^{(n)}+m\delta_{b}^{(m)} &=&\frac{(\Delta_{a}^{(n)}+\Delta_{b}^{(m)})\pm \sqrt{\mu^{2}(\Delta_{a}^{(n) }+\Delta_{b}^{(m)})^{2}+4(\mu^{2}-1)\Omega_{L}^{2}}}{(\mu^{2}-1)},
\end{eqnarray}
where $\text{\textquotedblleft }+\text{\textquotedblright}$ and $\text{\textquotedblleft }-\text{\textquotedblright}$ correspond to the higher-order transitions $|0\rangle_{a}|0\rangle_{b}|+\rangle \leftrightarrow |\mu n\rangle_{a}|\mu m\rangle_{b} |-\rangle$ and $|0\rangle_{a}|0\rangle_{b}|-\rangle \leftrightarrow |\mu n\rangle_{a}|\mu m\rangle_{b}|+\rangle$, respectively.

In the nondegenerate multiphoton JC model, we have identified the two resonant transition processes that lead to the ideal super-Rabi oscillation. In the following, we focus on the resonant transition $|k\rangle_{a}|l\rangle_{b}|+\rangle\leftrightarrow |k+n\rangle_{a}|l+m\rangle_{b}|-\rangle$, and assume that the initial state of the system is $|0\rangle_{a}|0\rangle_{b}|+\rangle$. In this case, the system can be approximated as an effective two-level system with the states $|0\rangle_{a}|0\rangle_{b}|+\rangle$ and $|n\rangle_{a}|m\rangle_{b}|-\rangle$. Therefore, the super-Rabi oscillations between the two states can be realized. By adiabatically eliminating the intermediate incoherent states and preserving only the coupling between the two involved states, the effective oscillating frequency of the super-Rabi oscillation can be obtained as (see Appendix)
\begin{equation}
	\Omega_\text{eff}^{(n,m)}=\frac{\sqrt{n!m!}gc_{+}^{2}(n\delta_{a}^{(n)}+m\delta_{b}^{(m)}+E_{+})E_{-}}{n!m!g^{2}c_{-}^{4}-(n\delta_{a}^{(n)}+m\delta_{b}^{(m)}+E_{+})E_{-}}. \label{Omega}
\end{equation}
The super-Rabi oscillation can be confirmed by checking the dynamic populations of the two states $|0\rangle_{a}|0\rangle_{b}|+\rangle$ and $|n\rangle_{a}|m\rangle_{b}|-\rangle$.

Figure \ref{Fig2} shows the state populations $P_{|0\rangle_{a}|0\rangle_{b}|+\rangle}$ and $P_{|n\rangle_{a}|m\rangle_{b}|-\rangle}$ ($n, m=1, 2$) as functions of the scaled evolution time $gt$. Here, the red solid and blue dashed curves represent the numerical results obtained using the Schr\"{o}dinger equation with Hamiltonian (\ref{HI}) and initial state $|0\rangle_{a}|0\rangle_{b}|+\rangle$, while the red points and the blue rectangles correspond to the analytical results based on the effective oscillating frequency $\Omega_{\text{eff}}^{(n,m)}$. It shows that a perfect super-Rabi oscillation can be achieved between the states $|0\rangle_{a}|0\rangle_{b}|+\rangle$ and $|n\rangle_{a}|m\rangle_{b}|-\rangle$ ($n, m=1, 2$). Moreover, the plots show that the analytical results match well with the numerical results.

\begin{figure}
	\center
	\includegraphics[width=0.49 \textwidth]{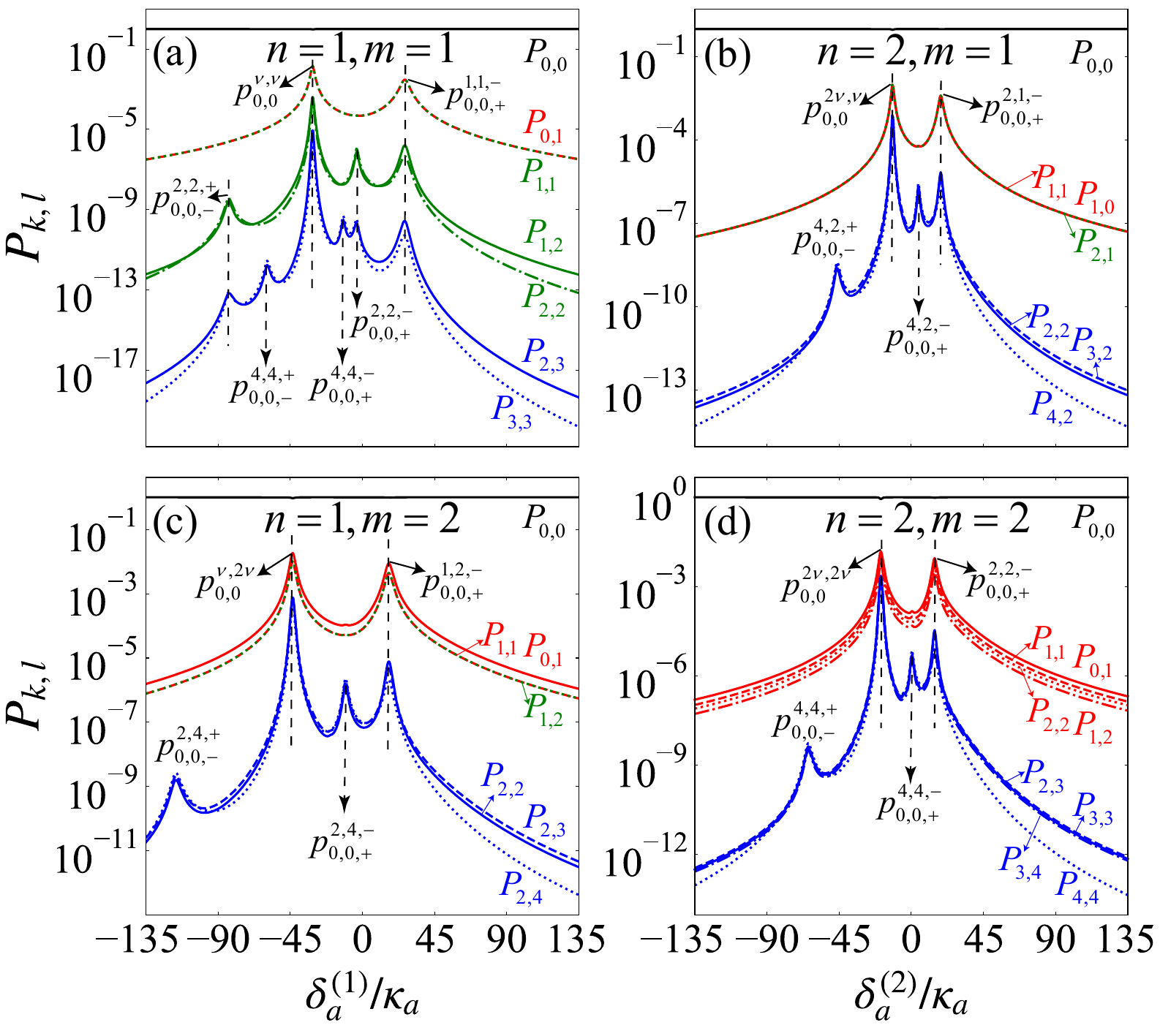}
	\caption{(Color online) The joint photon number distributions $P_{k,l}$ as functions of the scaled detuning $\delta_{a}^{(n)}/\kappa_{a}$ for a) the nondegenerate two-photon JC model ($n=1, m=1$) at $\Delta_{a}^{(1)}/\kappa_{a}=-16.5$, $\Delta_{b}^{(1)}/\kappa_{a}=-21$, and $\Omega_{L}/\kappa_{a}=27$; b) the nondegenerate three-photon JC model ($n=2, m=1$) at $\Delta_{a}^{(2)}/\kappa_{a}=-34.5$, $\Delta_{b}^{(1)}/\kappa_{a}=-21$, and $\Omega_{L}/\kappa_{a}=30$; c) the nondegenerate three-photon JC model ($n=1, m=2$) at $\Delta_{a}^{(1)}/\kappa_{a}=-16.5$, $\Delta_{b}^{(2)}/\kappa_{a}=-43.5$, and $\Omega_{L}/\kappa_{a}=30$; and d) the nondegenerate four-photon JC model ($n=2, m=2$) at $\Delta_{a}^{(2)}/\kappa_{a}=-34.5$, $\Delta_{b}^{(2)}/\kappa_{a}=-43.5$, and $\Omega_{L}/\kappa_{a}=33$. The detuning is taken as $\delta_{b}^{(m)}=-{[(\Delta_{a}^{(n)}-3\Delta_{b}^{(m)})(\Delta_{a}^{(n)}+\Delta_{b}^{(m)})-4\Omega_{L}^{2}}]/[4m(\Delta_{a}^{(n)}+\Delta_{b}^{(m)})]$. Other parameters used are $\gamma/\kappa_{a}=0.1$, $g/\kappa_{a}=0.3$, and $\kappa_{b}/\kappa_{a}=1$.}
	\label{Fig3}
\end{figure}

\subsection{The joint probability of the two-mode photon states}\label{subsectionB}
In the above subsection, we elaborate on the physical mechanism of the super-Rabi oscillations generated in the nondegenerate multiphoton JC model. In this subsection, we focus on the evolution of the system in the presence of dissipations. For typical quantum optical systems, the dissipations of the system can be described under the Markovian open-system framework. For the present system, we assume that the TLS and the two cavity-modes are connected with three separate vacuum baths, then the evolution of the system is governed by the quantum master equation
\begin{equation}
	\dot{\hat{\rho}}=-i[\hat{H}_{I},\hat{\rho}]+\kappa_{a}\mathcal{L}[\hat{a}]\hat{\rho}+\kappa_{b}\mathcal{L}[\hat{b}]\hat{\rho}+\gamma\mathcal{L}[\hat{\sigma}_{-}]\hat{\rho},\label{eq:21}
\end{equation}
where the Hamiltonian $\hat{H}_{I}$ is given by Equation ({\ref{HI}}), $\kappa_{a}$ ($\kappa_{b}$) is the decay rate of the optical mode $a$ ($b$), and $\gamma$ is the decay rate of the TLS. The standard Lindblad super-operators are defined by $\mathcal{L}[\hat{o}]\hat{\rho}=(2\hat{o}\hat{\rho}\hat{o}^{\dagger}-\hat{\rho}\hat{o}^{\dagger}\hat{o}-\hat{o}^{\dagger}\hat{o}\hat{\rho})/2$ for $\hat{o}=\hat{a}, \hat{b}$, and $\hat{\sigma}_{-}$. By solving Equation ({\ref{eq:21}}) with the Qutip toolbox,\textsuperscript{\cite{Johansson2012,Johansson2013}} we obtain the steady-state density operator $\hat{\rho}_{ss}$ of the system, i.e., the steady-state solution of Equation ({\ref{eq:21}}).

In order to study the two-mode correlated multiphoton bundle emission, we calculate the joint two-mode photon number distributions $P_{k,l}=\text{Tr}(|k\rangle_{a}|l\rangle_{b}\,_{a}\langle k|\,_{b}\langle l|\hat{\rho}_{ss})$, which describes the joint probability for finding $k$ photons in mode $a$ and $l$ photons in mode $b$. In Figure \ref{Fig3}a, we show the photon number distributions $P_{k,l}$~($k, l=0$-$3$) versus the scaled detuning $\delta_{a}^{(1)}/\kappa_{a}$ for the case ($n=1, m=1$). It can be observed that both the single-photon distribution $P_{0,1}$ and two-photon distribution $P_{1,1}$ have two distinct peaks located at $\delta_{a}^{(1)}/\kappa_{a}=-31.065$ and $\delta_{a}^{(1)}/\kappa_{a}=26.565$. By analyzing the eigenenergy levels (as shown in Figure \ref{Fig1}c) of the Hamiltonian $\hat{H}_{0}^{\prime}$ given in Equation (\ref{H_{0}}), we find that the peak located at $\delta_{a}^{(1)}/\kappa_{a}=-31.065$ corresponds to all these degenerate $2\nu$-photon ($\nu=1, \mu$) resonant transitions $|0\rangle_{a}|0\rangle_{b}|+\rangle\leftrightarrow |\nu\rangle_{a}|\nu\rangle_{b}|+\rangle$ and $|0\rangle_{a}|0\rangle_{b}|-\rangle\leftrightarrow |\nu\rangle_{a}|\nu\rangle_{b}|-\rangle$, while the peak located at $\delta_{a}^{(1)}/\kappa_{a}=26.565$ corresponds to the two-photon resonant transition $|0\rangle_{a}|0\rangle_{b}|+\rangle\leftrightarrow |1\rangle_{a}|1\rangle_{b}|-\rangle$. To be more clearer, we mark the two peaks as $p_{0,0}^{\nu,\nu}$ ($\nu=1, \mu$) and $p_{0,0,+}^{1,1,-}$. In addition, the three-photon distribution $P_{1,2}$ and the four-photon distribution $P_{2,2}$ have four peaks, two of these peaks are located at the same positions as the peaks of $P_{0,1}$ and $P_{1,1}$, corresponding to the mentioned $2\nu$-photon and two-photon resonant transitions. The other two peaks ($p_{0,0;+}^{2,2,-}$ and $p_{0,0,-}^{2,2,+}$) are located at $\delta_{a}^{(1)}/\kappa_{a}\approx-3.60$ and $\delta_{a}^{(1)}/\kappa_{a}\approx-83.53$, corresponding to the four-photon resonant transitions $|0\rangle_{a}|0\rangle_{b}|+\rangle\leftrightarrow |2\rangle_{a}|2\rangle_{b}|-\rangle$ and $|0\rangle_{a}|0\rangle_{b}|-\rangle\leftrightarrow |2\rangle_{a}|2\rangle_{b}|+\rangle$, respectively. Similarly, the five-photon distribution $P_{2,3}$ and the six-photon distribution $P_{3,3}$ have six bunching  peaks, four of which have the same positions as the peaks of $P_{1,2}$ and $P_{2,2}$. The other two peaks ($p_{0,0,+}^{3,3,-}$ and $p_{0,0,-}^{3,3,+}$) are located at $\delta_{a}^{(1)}/\kappa_{a}\approx-12.04$ and $\delta_{a}^{(1)}/\kappa_{a}\approx-59.46$, corresponding to the six-photon resonant transitions $|0\rangle_{a}|0\rangle_{b}|+\rangle\leftrightarrow |3\rangle_{a}|3\rangle_{b}|-\rangle$ and $|0\rangle_{a}|0\rangle_{b}|-\rangle\leftrightarrow |3\rangle_{a}|3\rangle_{b}|+\rangle$, respectively.

In Figure \ref{Fig3}b, we show the joint photon number distributions $P_{k,l}$~($k=0$-$4, l=0$-$2$) versus the scaled detuning $\delta_{a}^{(2)}/\kappa_{a}$ for the case ($n=2, m=1$). It can be observed that the distributions $P_{1,0}$, $P_{1,1}$, and $P_{2,1}$ have a peak ($p_{0,0}^{2\nu,\nu}$) at $\delta_{a}^{(2)}/\kappa_{a}\approx-11.67$, which corresponds to all these degenerate $3\nu$-photon ($\nu=1, \mu$) resonant transitions $|0\rangle_{a}|0\rangle_{b}|+\rangle\leftrightarrow |2\nu\rangle_{a}|\nu\rangle_{b}|+\rangle$ and $|0\rangle_{a}|0\rangle_{b}|-\rangle\leftrightarrow |2\nu\rangle_{a}|\nu\rangle_{b}|-\rangle$. Meanwhile, there is a peak ($p_{0,0,+}^{2,1,-}$) at $\delta_{a}^{(2)}/\kappa_{a}\approx18.42$, which corresponds to the three-photon resonant transition $|0\rangle_{a}|0\rangle_{b}|+\rangle\leftrightarrow |2\rangle_{a}|1\rangle_{b}|-\rangle$. In addition, the distributions $P_{2,2}$, $P_{3,2}$, and $P_{4,2}$ have four bunching peaks, two of these peaks are located at  the same positions as the peaks of $P_{2,1}$, corresponding to the mentioned $3\nu$-photon resonant transition and the three-photon resonant transition. The positions of the other two peaks ($p_{0,0,+}^{4,2,-}$ and $p_{0,0,-}^{4,2,+}$) are determined by the six-photon resonant transitions $|0\rangle_{a}|0\rangle_{b}|+\rangle\leftrightarrow |4\rangle_{a}|2\rangle_{b}|-\rangle$ and $|0\rangle_{a}|0\rangle_{b}|-\rangle\leftrightarrow |4\rangle_{a}|2\rangle_{b}|+\rangle$. 

In Figure \ref{Fig3}c, we show the joint photon number distributions $P_{k,l}$~($k=0$-$2, l=0$-$4$) versus the scaled detuning $\delta_{a}^{(1)}/\kappa_{a}$ for the case ($n=1, m=2$). These four peaks ($p_{0,0}^{\nu,2\nu}$, $p_{0,0,+}^{1,2,-}$, $p_{0,0,+}^{2,4,-}$, and $p_{0,0,-}^{2,4,+}$) located at $\delta_{a}^{(1)}/\kappa_{a}=-43.5$, $\delta_{a}^{(1)}/\kappa_{a}=16.5$, $\delta_{a}^{(1)}/\kappa_{a}\approx-10.58$, and $\delta_{a}^{(1)}/\kappa_{a}\approx-116.42$ correspond to $3\nu$-photon ($\nu=1, \mu$) resonant transitions $|0\rangle_{a}|0\rangle_{b}|+\rangle\leftrightarrow |\nu\rangle_{a}|2\nu\rangle_{b}|+\rangle$ ($|0\rangle_{a}|0\rangle_{b}|-\rangle\leftrightarrow |\nu\rangle_{a}|2\nu\rangle_{b}|-\rangle$), three-photon resonant transition $|0\rangle_{a}|0\rangle_{b}|+\rangle\leftrightarrow |1\rangle_{a}|2\rangle_{b}|-\rangle$, six-photon resonant transition $|0\rangle_{a}|0\rangle_{b}|+\rangle\leftrightarrow |2\rangle_{a}|4\rangle_{b}|-\rangle$, and six-photon resonant transition $|0\rangle_{a}|0\rangle_{b}|-\rangle\leftrightarrow |2\rangle_{a}|4\rangle_{b}|+\rangle$, respectively. In Figure \ref{Fig3}d, we show the joint photon number distributions $P_{k,l}$~($k, l=0$-$4$) versus the scaled detuning $\delta_{a}^{(2)}/\kappa_{a}$ for the case ($n=2, m=2$). Similarly, these four peaks ($p_{0,0}^{2\nu,2\nu}$, $p_{0,0,+}^{2,2,-}$, $p_{0,0.+}^{4,4,-}$, and $p_{0,0,-}^{4,4,+}$) located at $\delta_{a}^{(2)}/\kappa_{a}\approx-18.98$, $\delta_{a}^{(2)}/\kappa_{a}\approx14.48$, $\delta_{a}^{(2)}/\kappa_{a}\approx0.25$, and $\delta_{a}^{(2)}/\kappa_{a}\approx-64.21$ correspond to $4\nu$-photon ($\nu=1, \mu$) resonant transitions $|0\rangle_{a}|0\rangle_{b}|+\rangle\leftrightarrow |2\nu\rangle_{a}|2\nu\rangle_{b}|+\rangle$ ($|0\rangle_{a}|0\rangle_{b}|-\rangle\leftrightarrow |2\nu\rangle_{a}|2\nu\rangle_{b}|-\rangle$), four-photon resonant transition $|0\rangle_{a}|0\rangle_{b}|+\rangle\leftrightarrow |2\rangle_{a}|2\rangle_{b}|-\rangle$, eight-photon resonant transition $|0\rangle_{a}|0\rangle_{b}|+\rangle\leftrightarrow |4\rangle_{a}|4\rangle_{b}|-\rangle$, and eight-photon resonant transition $|0\rangle_{a}|0\rangle_{b}|-\rangle\leftrightarrow |4\rangle_{a}|4\rangle_{b}|+\rangle$, respectively.

\begin{figure}
	\center
	\includegraphics[width=0.49 \textwidth]{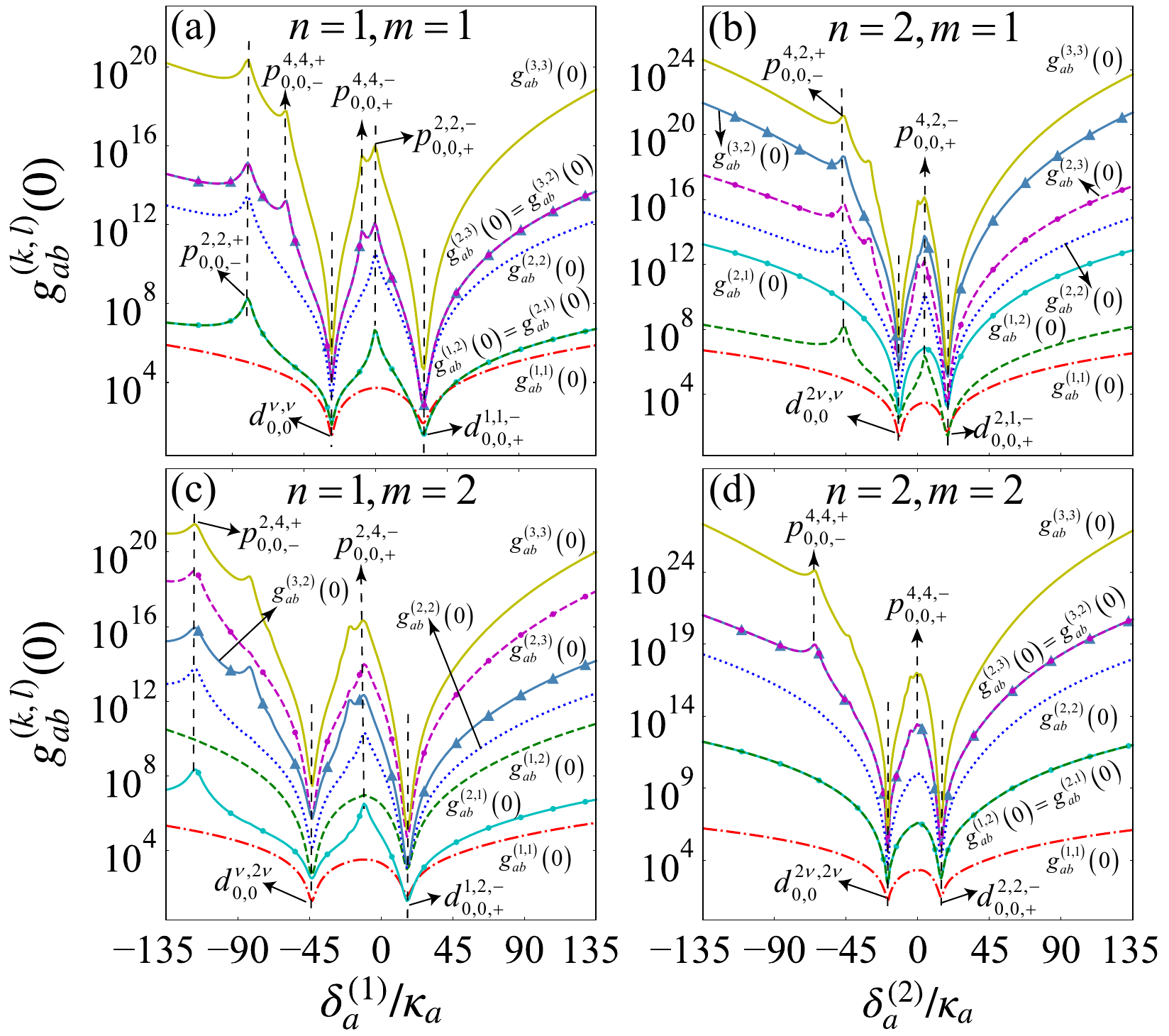}
	\caption{(Color online) The equal-time high-order correlation functions $\mathrm{g}^{(k,l)}_{ab}(0)$ as functions of the scaled detuning $\delta_{a}^{(n)}/\kappa_{a}$ for a) the nondegenerate two-photon JC model ($n=1, m=1$) at $\Delta_{a}^{(1)}/\kappa_{a}=-16.5$, $\Delta_{b}^{(1)}/\kappa_{a}=-21$, and $\Omega_{L}/\kappa_{a}=27$; b) the nondegenerate three-photon JC model ($n=2, m=1$) at $\Delta_{a}^{(2)}/\kappa_{a}=-34.5$, $\Delta_{b}^{(1)}/\kappa_{a}=-21$, and $\Omega_{L}/\kappa_{a}=30$; c) the nondegenerate three-photon JC model ($n=1, m=2$) at $\Delta_{a}^{(1)}/\kappa_{a}=-16.5$, $\Delta_{b}^{(2)}/\kappa_{a}=-43.5$, and $\Omega_{L}/\kappa_{a}=30$; and d) the nondegenerate four-photon JC model ($n=2, m=2$) at $\Delta_{a}^{(2)}/\kappa_{a}=-34.5$, $\Delta_{b}^{(2)}/\kappa_{a}=-43.5$, and $\Omega_{L}/\kappa_{a}=33$. The detuning is taken as $\delta_{b}^{(m)}=-{[(\Delta_{a}^{(n)}-3\Delta_{b}^{(m)})(\Delta_{a}^{(n)}+\Delta_{b}^{(m)})-4\Omega_{L}^{2}}]/[4m(\Delta_{a}^{(n)}+\Delta_{b}^{(m)})]$. Other parameters used are $\gamma/\kappa_{a}=0.1$, $g/\kappa_{a}=0.3$, and $\kappa_{b}/\kappa_{a}=1$.}
	\label{Fig4}
\end{figure}

\subsection{Two-mode equal-time high-order correlation function}\label{subsectionC}
To further describe the quantum statistics of the nondegenerate multiphoton JC model, we analyze the equal-time high-order correlation function of the two cavity modes\textsuperscript{\cite{Stassi2013,Ou2017}}
\begin{equation}
	\mathrm{g}^{(k,l)}_{ab}(0)=\frac{\textrm{Tr}(\hat{a}^{\dagger k}\hat{b}^{\dagger l}\hat{b}^{l}\hat{a}^{k}\hat{\rho}_{ss})}{[\textrm{Tr}(\hat{a}^{\dagger}\hat{a}\hat{\rho}_{ss})]^{k}[\textrm{Tr}(\hat{b}^{\dagger}\hat{b}\hat{\rho}_{ss})]^{l}},
\end{equation}
where $\hat{\rho}_{ss}$ is the steady-state density operator of the system. In Figure \ref{Fig4}a, we plot the equal-time high-order correlation functions ${g}^{(k,l)}_{ab}(0)$ ($k, l=1$-$3$) as functions of the scaled detuning $\delta_{a}^{(1)}/\kappa_{a}$ for the case ($n=1, m=1$). It can be seen that the values of ${g}^{(1,1)}_{ab}(0)$ is greater than 1 in all parameter ranges, indicating a strong correlation among the emitted photons.\textsuperscript{\cite{Gou2022,Ma2022}} Moreover, we observe sharp dips ($d_{0,0}^{\nu,\nu}$ and $d_{0,0,+}^{1,1,-}$) located at $\delta_{a}^{(1)}/\kappa_{a}=-31.065$ and $\delta_{a}^{(1)}/\kappa_{a}=26.565$ for the correlation functions ${g}^{(k,l)}_{ab}(0)$ ($k, l=1$-$3$). The two dips correspond to the $2\nu$-photon ($\nu=1, \mu$) resonant transitions $|0\rangle_{a}|0\rangle_{b}|+\rangle\leftrightarrow |\nu\rangle_{a}|\nu\rangle_{b}|+\rangle$ ($|0\rangle_{a}|0\rangle_{b}|-\rangle\leftrightarrow |\nu\rangle_{a}|\nu\rangle_{b}|-\rangle$) and the two-photon resonant transition $|0\rangle_{a}|0\rangle_{b}|+\rangle\leftrightarrow |1\rangle_{a}|1\rangle_{b}|-\rangle$, respectively. In these two cases, the system enters the regime of two-photon bundle emission. Furthermore, ${g}^{(1,2)}_{ab}(0)$, ${g}^{(2,1)}_{ab}(0)$, and ${g}^{(2,2)}_{ab}(0)$ exhibit two peaks ($p_{0,0,+}^{2,2,-}$ and $p_{0,0,-}^{2,2,+}$) at $\delta_{a}^{(1)}/\kappa_{a}\approx-3.60$ and $\delta_{a}^{(1)}/\kappa_{a}\approx-83.53$. The two peaks correspond to the four-photon resonant transitions $|0\rangle_{a}|0\rangle_{b}|+\rangle\leftrightarrow |2\rangle_{a}|2\rangle_{b}|-\rangle$ and $|0\rangle_{a}|0\rangle_{b}|-\rangle\leftrightarrow |2\rangle_{a}|2\rangle_{b}|+\rangle$, respectively. For ${g}^{(2,3)}_{ab}(0)$, ${g}^{(3,2)}_{ab}(0)$, and ${g}^{(3,3)}_{ab}(0)$, there are two bunching peaks located at the same positions as those of the peaks of ${g}^{(1,2)}_{ab}(0)$ and ${g}^{(2,2)}_{ab}(0)$, as well as two additional peaks ($p_{0,0,+}^{3,3,-}$ and $p_{0,0,-}^{3,3,+}$) located at $\delta_{a}^{(1)}/\kappa_{a}\approx-12.04$ and $\delta_{a}^{(1)}/\kappa_{a}\approx-59.46$, corresponding to the six-photon resonant transitions $|0\rangle_{a}|0\rangle_{b}|+\rangle\leftrightarrow |3\rangle_{a}|3\rangle_{b}|-\rangle$ and $|0\rangle_{a}|0\rangle_{b}|-\rangle\leftrightarrow |3\rangle_{a}|3\rangle_{b}|+\rangle$, respectively.

In Figure \ref{Fig4}b, we show the equal-time high-order correlation functions ${g}^{(k,l)}_{ab}(0)$ ($k, l=1$-$3$) for the case ($n=2, m=1$) as functions of the scaled detuning $\delta_{a}^{(2)}/\kappa_{a}$. Similar to Figure \ref{Fig4}a, all the values of the correlation function ${g}^{(1,1)}_{ab}(0)$ is greater than 1, indicating a strong correlation between the photons. Moreover, the correlation functions ${g}^{(k,l)}_{ab}(0)$ ($k, l=1$-$3$) show two sharp dips ($d_{0,0}^{2\nu,\nu}$ and $d_{0,0,+}^{2,1,-}$) at $\delta_{a}^{(2)}/\kappa_{a}\approx-11.67$ and $\delta_{a}^{(2)}/\kappa_{a}\approx18.42$. The locations of these two dips are determined by the $3\nu$-photon ($\nu=1, \mu$) resonant transitions $|0\rangle_{a}|0\rangle_{b}|+\rangle\leftrightarrow |2\nu\rangle_{a}|\nu\rangle_{b}|+\rangle$ ($|0\rangle_{a}|0\rangle_{b}|-\rangle\leftrightarrow |2\nu\rangle_{a}|\nu\rangle_{b}|-\rangle$) and the three-photon resonant transition $|0\rangle_{a}|0\rangle_{b}|+\rangle\leftrightarrow |2\rangle_{a}|1\rangle_{b}|-\rangle$. Under these two conditions, the system enters the regime of three-photon bundle emission. In addition, ${g}^{(1,2)}_{ab}(0)$, ${g}^{(2,2)}_{ab}(0)$, and ${g}^{(3,2)}_{ab}(0)$ exhibit two peaks ($p_{0,0,+}^{4,2,-}$ and $p_{0,0,-}^{4,2,+}$) at $\delta_{a}^{(2)}/\kappa_{a}\approx4.42$ and $\delta_{a}^{(2)}/\kappa_{a}\approx-46.26$, corresponding to the six-photon resonant transitions $|0\rangle_{a}|0\rangle_{b}|+\rangle\leftrightarrow |4\rangle_{a}|2\rangle_{b}|-\rangle$ and $|0\rangle_{a}|0\rangle_{b}|-\rangle\leftrightarrow |4\rangle_{a}|2\rangle_{b}|+\rangle$, respectively.

In Figure \ref{Fig4}c, we show the equal-time high-order correlation functions ${g}^{(k,l)}_{ab}(0)$ ($k, l=1$-$3$) for the case ($n=1, m=2$) as functions of the scaled detuning $\delta_{a}^{(1)}/\kappa_{a}$. It can be seen that the values of ${g}^{(1,1)}_{ab}(0)$ is greater than 1 in all parameter ranges, indicating a strong correlation among the emitted photons. Meanwhile, two sharp dips ($d_{0,0}^{\nu,2\nu}$ and $d_{0,0,+}^{1,2,-}$) appear at $\delta_{a}^{(1)}/\kappa_{a}=-43.5$ and $\delta_{a}^{(2)}/\kappa_{a}=16.5$. These two dips correspond to the $3\nu$-photon ($\nu=1, \mu$) resonant transitions $|0\rangle_{a}|0\rangle_{b}|+\rangle\leftrightarrow |\nu\rangle_{a}|2\nu\rangle_{b}|+\rangle$ ($|0\rangle_{a}|0\rangle_{b}|-\rangle\leftrightarrow |\nu\rangle_{a}|2\nu\rangle_{b}|-\rangle$) and the three-photon resonant transition $|0\rangle_{a}|0\rangle_{b}|+\rangle\leftrightarrow |1\rangle_{a}|2\rangle_{b}|-\rangle$. This indicates that the system enters the regime of three-photon bundle emission. The correlation functions ${g}^{(2,1)}_{ab}(0)$, ${g}^{(2,2)}_{ab}(0)$, and ${g}^{(2,3)}_{ab}(0)$ exhibit two peaks ($p_{0,0,+}^{2,4,-}$ and $p_{0,0,-}^{2,4,+}$) at $\delta_{a}^{(1)}/\kappa_{a}\approx-10.58$ and $\delta_{a}^{(1)}/\kappa_{a}\approx-116.42$, corresponding to the six-photon resonant transitions $|0\rangle_{a}|0\rangle_{b}|+\rangle\leftrightarrow |2\rangle_{a}|4\rangle_{b}|-\rangle$ and $|0\rangle_{a}|0\rangle_{b}|-\rangle\leftrightarrow |2\rangle_{a}|4\rangle_{b}|+\rangle$, respectively.

In Figure \ref{Fig4}d, we show the equal-time high-order correlation functions ${g}^{(k,l)}_{ab}(0)$ ($k, l=1$-$3$) for the case ($n=2, m=2$) as functions of the scaled detuning $\delta_{a}^{(2)}/\kappa_{a}$. The correlation function ${g}^{(1,1)}_{ab}(0)$ is greater than 1 in all parameter ranges, indicating a strong correlation among the emitted photons. Additionally, two sharp dips ($d_{0,0}^{2\nu,2\nu}$ and $d_{0,0,+}^{2,2,-}$) appear at $\delta_{a}^{(2)}/\kappa_{a}\approx-18.98$ and $\delta_{a}^{(2)}/\kappa_{a}\approx14.48$. These two dips correspond to the $4\nu$-photon ($\nu=1, \mu$) resonant transitions $|0\rangle_{a}|0\rangle_{b}|+\rangle\leftrightarrow |2\nu\rangle_{a}|2\nu\rangle_{b}|+\rangle$ ($|0\rangle_{a}|0\rangle_{b}|-\rangle\leftrightarrow |2\nu\rangle_{a}|2\nu\rangle_{b}|-\rangle$) and the four-photon resonant transition $|0\rangle_{a}|0\rangle_{b}|+\rangle\leftrightarrow |2\rangle_{a}|2\rangle_{b}|-\rangle$. This indicates that the system enters the regime of four-photon bundle emission. The correlation function ${g}^{(2,3)}_{ab}(0)$, ${g}^{(3,2)}_{ab}(0)$, and ${g}^{(3,3)}_{ab}(0)$ exhibits two peaks ($p_{0,0,+}^{4,4,-}$ and $p_{0,0,-}^{4,4,+}$) at $\delta_{a}^{(2)}/\kappa_{a}\approx0.25$ and $\delta_{a}^{(2)}/\kappa_{a}\approx-64.21$, corresponding to the eight-photon resonant transitions $|0\rangle_{a}|0\rangle_{b}|+\rangle\leftrightarrow |4\rangle_{a}|4\rangle_{b}|-\rangle$ and $|0\rangle_{a}|0\rangle_{b}|-\rangle\leftrightarrow |4\rangle_{a}|4\rangle_{b}|+\rangle$, respectively.
\begin{figure}
	\center
	\includegraphics[width=0.49 \textwidth]{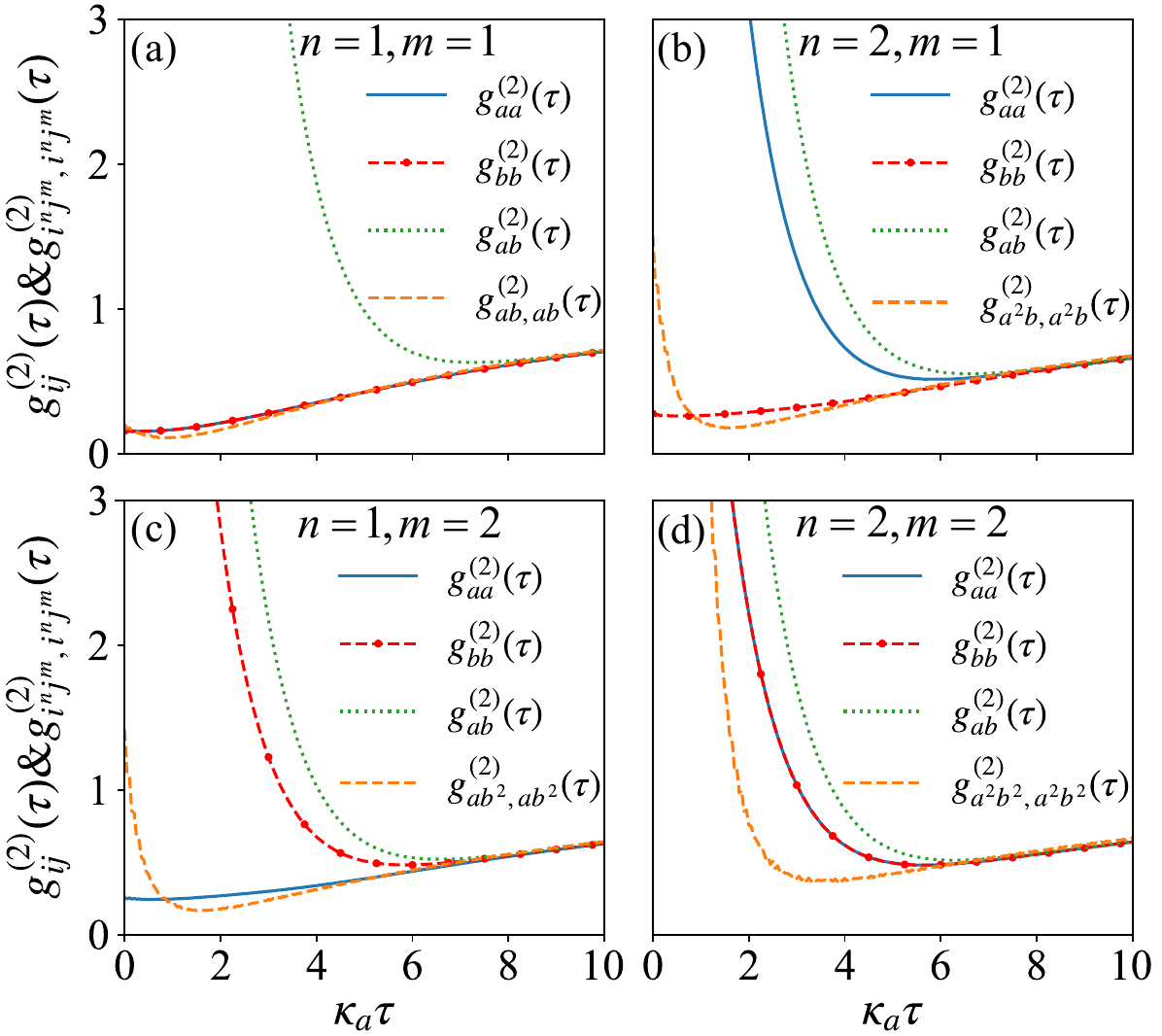}
	\caption{(Color online) The time-delayed second-order correlation functions ${g}_{ij}^{(2)}(\tau)$ and $\mathrm{g}_{i^{n}j^{m},i^{n}j^{m}}^{(2)}(\tau)$ for $\{i, j\}=\{a, b\}$ as functions of time $\kappa_{a}\tau$ in various physical models under proper parameters:a) the nondegenerate two-photon JC model ($n=1, m=1$) at $\Delta_{a}^{(1)}/\kappa_{a}=-16.5$, $\Delta_{b}^{(1)}/\kappa_{a}=-21$, and $\Omega_{L}/\kappa_{a}=27$; b) the nondegenerate three-photon JC model ($n=2, m=1$) at $\Delta_{a}^{(2)}/\kappa_{a}=-34.5$, $\Delta_{b}^{(1)}/\kappa_{a}=-21$, and $\Omega_{L}/\kappa_{a}=30$; c) the nondegenerate three-photon JC model ($n=1, m=2$) at $\Delta_{a}^{(1)}/\kappa_{a}=-16.5$, $\Delta_{b}^{(2)}/\kappa_{a}=-43.5$, and $\Omega_{L}/\kappa_{a}=30$; and d) the nondegenerate four-photon JC model ($n=2, m=2$) at $\Delta_{a}^{(2)}/\kappa_{a}=-34.5$, $\Delta_{b}^{(2)}/\kappa_{a}=-43.5$, and $\Omega_{L}/\kappa_{a}=33$. The detunings are taken as $\delta_{a}^{(n)}=-{[(\Delta_{b}^{(m)}-3\Delta_{a}^{(n)})(\Delta_{a}^{(n)}+\Delta_{b}^{(m)})-4\Omega_{L}^{2}}]/[4n(\Delta_{a}^{(n)}+\Delta_{b}^{(m)})]$ and $\delta_{b}^{(m)}=-{[(\Delta_{a}^{(n)}-3\Delta_{b}^{(m)})(\Delta_{a}^{(n)}+\Delta_{b}^{(m)})-4\Omega_{L}^{2}}]/[4m(\Delta_{a}^{(n)}+\Delta_{b}^{(m)})]$. Other parameters used are $\gamma/\kappa_{a}=0.1$, $g/\kappa_{a}=0.3$, and $\kappa_{b}/\kappa_{a}=1$.}
	\label{Fig5}
\end{figure}

\subsection{Time-delayed second-order correlation function}\label{subsectionD}
To better investigate the quantum statistics of the nondegenerate multiphoton JC model, we further analyze the time-delayed second-order correlation function\textsuperscript{\cite{Ou2017,Mandel1995}}
\begin{equation}
	\mathrm{g}_{ij}^{(2)}(\tau)=\frac{\langle \hat{i}^{\dagger }(0)\hat{j}^{\dagger }(\tau)\hat{j}(\tau)\hat{i}(0)\rangle}{\langle (\hat{i}^{\dagger }\hat{i})(0)\rangle\langle (\hat{j}^{\dagger }\hat{j})(\tau)\rangle},\label{gn}
\end{equation}
where $\{i, j\}=\{a, b\}$ and $\tau$ is delayed time. When $i=j$, the correlation function (\ref{gn}) is reduced to the standard time-delayed second-order correlation function that measures the correlation between the photons emitted by a single cavity. When $i\neq j$, in contrast, it characterizes the correlation function of the emitted photons between different optical cavities. Moreover, to accurately describe the two-mode correlated multiphoton bundle emission, we calculate the generalized time-delayed second-order correlation function of the multiphoton bundle\textsuperscript{\cite{Ma2022}}
\begin{equation}
	\mathrm{g}_{i^{N}j^{M},i^{N}j^{M}}^{(2)}(\tau)=\frac{G_{i^{N}j^{M},i^{N}j^{M}}(0,\tau)}{G_{i^{N}j^{M},i^{N}j^{M}}(0)G_{i^{N}j^{M},i^{N}j^{M}}(\tau)},\label{gn2}
\end{equation}
where 
\begin{subequations}
	\begin{align}
		G_{i^{N}j^{M},i^{N}j^{M}}(0,\tau) =&{\langle\hat{i}^{\dagger N}(0)\hat{j}^{\dagger M}(0)\hat{i}^{\dagger N}(\tau)\hat{j}^{\dagger M}(\tau)} \notag \\
		&\times{\hat{i}^{N}(\tau)\hat{j}^{M}(\tau)\hat{i}^{N}(0)\hat{j}^{M}(0)\rangle} , \\
		G_{i^{N}j^{M},i^{N}j^{M}}(0) =&\langle(\hat{i}^{\dagger N}\hat{j}^{\dagger M}\hat{i}^{N}\hat{j}^{M})(0)\rangle ,  \\
		G_{i^{N}j^{M},i^{N}j^{M}}(\tau) =&\langle(\hat{i}^{\dagger N}\hat{j}^{\dagger M}\hat{i}^{N}\hat{j}^{M})(\tau)\rangle
	\end{align}
\end{subequations}
for $\{i, j\}=\{a, b\}$ and $\tau$ is delayed time. In a realistic case, it will take some time to finish the emission of multiple photons. Therefore, the correlation function $\mathrm{g}_{i^{N}j^{M},i^{N}j^{M}}^{(2)}(\tau)$ is ill-defined in a short temporal window of width $\tau^{[N,M]}_{\text{min}}=\sum_{k=1}^{N}{(1/k\kappa_{a})}+\sum_{l=1}^{M}{(1/l\kappa_{b})}$,\textsuperscript{\cite{Gou2022,Jiang2022,Bin2020}} where $\tau^{[N,M]}_{\text{min}}$ is approximately regarded as the time duration corresponding to the multiphoton emission. As a result, the $\mathrm{g}_{i^{N}j^{M},i^{N}j^{M}}^{(2)}(\tau^{[N,M]}_{\text{min}})$ can be approximately regarded as the zero-time correlation function. The correlation function $\mathrm{g}_{i^{N}j^{M},i^{N}j^{M}}^{(2)}(\tau)$ can capture the fundamental dynamics of multiphoton bundle emission and characterize the statistical properties of the multiphoton bundle. The inequality $\mathrm{g}_{i^{N}j^{M},i^{N}j^{M}}^{(2)}(\tau^{[N,M]}_{\text{min}})<\mathrm{g}_{i^{N}j^{M},i^{N}j^{M}}^{(2)}(\tau)$ [$\mathrm{g}_{i^{N}j^{M},i^{N}j^{M}}^{(2)}(\tau^{[N,M]}_{\text{min}})>\mathrm{g}_{i^{N}j^{M},i^{N}j^{M}}^{(2)}(\tau)$] for $\tau\geq\tau^{[N,M]}_{\text{min}}$ indicates the anti-bunching (bunching) effect of the two-mode correlated multiphoton bundles.
\begin{figure*}[t]
	\centering	
	\includegraphics[width=0.99 \textwidth]{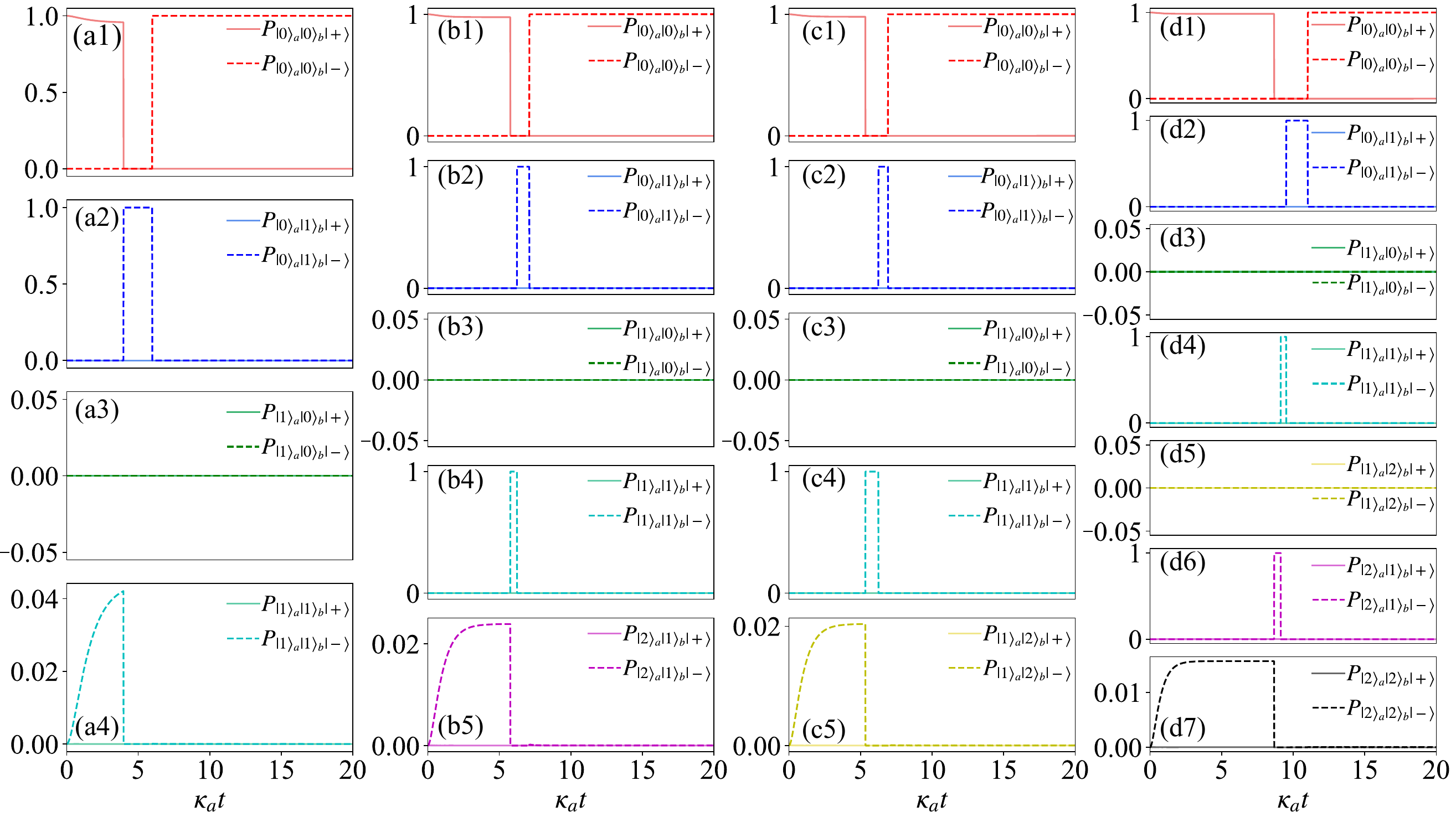}
	\caption{(Color online) A part of the quantum trajectory of the state populations $P_{|k\rangle_{a}|l\rangle_{b}|\pm\rangle}$ ($k, l=0, 1$, and $2$) for a1-a4) the nondegenerate two-photon JC model ($n=1, m=1$) at $\Delta_{a}^{(1)}/\kappa_{a}=-16.5$, $\Delta_{b}^{(1)}/\kappa_{a}=-21$, and $\Omega_{L}/\kappa_{a}=27$; b1-b5) the nondegenerate three-photon JC model ($n=2, m=1$) at $\Delta_{a}^{(2)}/\kappa_{a}=-34.5$, $\Delta_{b}^{(1)}/g=-21$, and $\Omega_{L}/\kappa_{a}=30$; c1-c5) the nondegenerate three-photon JC model ($n=1, m=2$) at $\Delta_{a}^{(1)}/\kappa_{a}=-16.5$, $\Delta_{b}^{(2)}/\kappa_{a}=-43.5$, and $\Omega_{L}/\kappa_{a}=30$; and d1-d7) the nondegenerate four-photon JC model ($n=2, m=2$) at $\Delta_{a}^{(2)}/\kappa_{a}=-34.5$, $\Delta_{b}^{(2)}/\kappa_{a}=-43.5$, and $\Omega_{L}/\kappa_{a}=33$. The detunings are taken as $\delta_{a}^{(n)}=-{[(\Delta_{b}^{(m)}-3\Delta_{a}^{(n)})(\Delta_{a}^{(n)}+\Delta_{b}^{(m)})-4\Omega_{L}^{2}}]/[4n(\Delta_{a}^{(n)}+\Delta_{b}^{(m)})]$ and $\delta_{b}^{(m)}=-{[(\Delta_{a}^{(n)}-3\Delta_{b}^{(m)})(\Delta_{a}^{(n)}+\Delta_{b}^{(m)})-4\Omega_{L}^{2}}]/[4m(\Delta_{a}^{(n)}+\Delta_{b}^{(m)})]$. Other parameters used are $\gamma/\kappa_{a}=0.1$, $g/\kappa_{a}=0.3$, and $\kappa_{b}/\kappa_{a}=1$.}
	\label{Fig6}
\end{figure*}

In Figure \ref{Fig5}a, we show the time-delayed second-order correlation functions ${g}_{ij}^{(2)}(\tau)$ for $\{i, j\}=\{a, b\}$ and ${g}_{ab,ab}^{(2)}(\tau)$ at detuning $\delta_{a}^{(1)}/\kappa_{a}=26.565$ for the nondegenerate two-photon JC model ($n=1, m=1$). The results clearly show that $\mathrm{g}_{aa}^{(2)}(0)<\mathrm{g}_{aa}^{(2)}(\tau)$ and $\mathrm{g}_{bb}^{(2)}(0)<\mathrm{g}_{bb}^{(2)}(\tau)$, which indicate the anti-bunding effect among the photons in the same cavity mode. In addition, it can be observed that $\mathrm{g}_{ab}^{(2)}(0)>\mathrm{g}_{ab}^{(2)}(\tau)$, which shows a strong correlation between the photons in different cavity modes. In the case where each of the two optical cavities emits a single photon at the same time, it is appropriate to describe the photon pair as a photon bundle. It can also be found that $\mathrm{g}_{ab,ab}^{(2)}(\tau^{[1,1]}_{\text{min}})<\mathrm{g}_{ab,ab}^{(2)}(\tau)$, which means that there is an anti-bunching effect between the photon bundles.

In Figure \ref{Fig5}b, we exhibit the photon statistics of the nondegenerate three-photon JC model ($n=2, m=1$). Here, we see the relation $\mathrm{g}_{aa}^{(2)}(0)>\mathrm{g}_{aa}^{(2)}(\tau)$ [$\mathrm{g}_{bb}^{(2)}(0)<\mathrm{g}_{bb}^{(2)}(\tau)$], indicating a  bunching (an anti-bunching) effect between the photons emitted in mode $a$ ($b$). It can also be observed that $\mathrm{g}_{ab}^{(2)}(0)>\mathrm{g}_{ab}^{(2)}(\tau)$, indicating that the photons emitted from the two optical cavities can form strongly correlated photon bundles. In addition, we find $\mathrm{g}_{a^{2}b,a^{2}b}^{(2)}(\tau^{[2,1]}_{\text{min}})<\mathrm{g}_{a^{2}b,a^{2}b}^{(2)}(\tau)$. This result indicates that there is a strong anti-bunching effect between the photon bundles formed by the photons emitted from the two cavity modes. 

In Figure \ref{Fig5}c, we show the time-delayed second-order correlation functions ${g}_{ab^{2},ab^{2}}^{(2)}(\tau)$ and ${g}_{ij}^{(2)}(\tau)$ for $\{i, j\}=\{a, b\}$ at $\delta_{a}^{(1)}/\kappa_{a}=16.5$ for the nondegenerate three-photon JC model ($n=1, m=2$). The anti-bunching and bunching effects can be seen between the photons emitted in modes $a$ and $b$, respectively. We also observe a strong anti-bunching effect between the photon bundles.

We also study the photon statistics in the nondegenerate four-photon JC model ($n=2, m=2$). In Figure \ref{Fig5}d, we show the time-delayed second-order correlation functions ${g}_{a^{2}b^{2},a^{2}b^{2}}^{(2)}(\tau)$ and ${g}_{ij}^{(2)}(\tau)$ for $\{i, j\}=\{a, b\}$ at detuning  $\delta_{a}^{(2)}/\kappa_{a}\approx14.48$. It can be found that $\mathrm{g}_{bb}^{(2)}(0)>\mathrm{g}_{bb}^{(2)}(\tau)$ and $\mathrm{g}_{aa}^{(2)}(0)>\mathrm{g}_{aa}^{(2)}(\tau)$, which means the bunching effect between photons of the same mode. In addition, it can be clearly found that $\mathrm{g}_{ab}^{(2)}(0)>\mathrm{g}_{ab}^{(2)}(\tau)$. The results show that the photons emitted by the two different modes can form strongly correlated photon bundles. We can also observe $\mathrm{g}_{a^{2}b^{2},a^{2}b^{2}}^{(2)}(\tau^{[2,2]}_{\text{min}})<\mathrm{g}_{a^{2}b^{2},a^{2}b^{2}}^{(2)}(\tau)$, which indicates an anti-bunching effect between the photon bundles.

\subsection{Quantum Monte-Carlo simulation}\label{subsectionE}
To exhibit the two-mode correlated multiphoton bundle emission process more clearly, we use quantum Monte-Carlo simulation to track the individual quantum trajectory of the system. In Figure \ref{Fig6}a1-a4, we show a part of quantum trajectory for the state populations $P_{|k\rangle_{a}|l\rangle_{b}|\pm\rangle}(t)$ ($k, l=0$ and $1$) when $\delta_{a}^{(1)}/\kappa_{a}=26.565$ for the nondegenerate two-photon JC model ($n=1, m=1$). Here we consider that the system is initially in the state $|0\rangle_{a}|0\rangle_{b}|+\rangle$. Under the two-photon resonant driving condition, the occupation probability $P_{|1\rangle_{a}|1\rangle_{b}|-\rangle}(t)$ of the two-photon state $|1\rangle_{a}|1\rangle_{b}|-\rangle$ is 0.044 at time $\kappa_{a}t\approx3.97$, while both the one-photon state populations $P_{|0\rangle_{a}|1\rangle_{b}|\pm\rangle}(t)$ and $P_{|1\rangle_{a}|0\rangle_{b}|\pm\rangle}(t)$ are 0. The dissipation of the optical field causes the emission of the first photon, and the wave function of the system collapses into the one-photon state ($|0\rangle_{a}|1\rangle_{b}|-\rangle$ or $|1\rangle_{a}|0\rangle_{b}|-\rangle$) with identical and almost unity probabilities. After that, the second photon is emitted within the cavity photon lifetime and the wave function of the system collapses into the zero-photon state $|0\rangle_{a}|0\rangle_{b}|-\rangle$, forming a two-mode correlated two-photon bundle emission. In the case of  $\Omega_{L}/\kappa_{a}=27$, $\Delta_{a}^{(1)}/\kappa_{a}=-16.5$, $\Delta_{b}^{(1)}/\kappa_{a}=-21$, and $\delta_{a}^{(1)}/\kappa_{a}=26.565$, there exists the relation $c_{-}\gg c_{+}$. Thus, the eigenstates $\vert-\rangle$ and $\vert+\rangle$ of the TLS can be approximately regarded as the bare states $\vert e\rangle$ and $\vert g\rangle$, respectively. This means that the system is approximately in the state $|0\rangle_{a}|0\rangle_{b}| e\rangle$ after the two-photon emission. Due to the dissipation of the TLS, the system will go back to the state $|0\rangle_{a}|0\rangle_{b}|g\rangle\approx|0\rangle_{a}|0\rangle_{b}|+\rangle$, reaching the beginning of the next two-photon bundle emission process. 

In Figure \ref{Fig6}b1-b5, we show a part of the quantum trajectory for the state populations $P_{|k\rangle_{a}|l\rangle_{b}|\pm\rangle}(t)$ ($k=0$-$2$ and $l=0, 1$) when $\delta_{a}^{(2)}/\kappa_{a}\approx18.42$ for the nondegenerate three-photon JC model ($n=2, m=1$). Under the condition of three-photon resonant driving (i.e., $\Omega_{L}/\kappa_{a}=30$, $\Delta_{a}^{(2)}/\kappa_{a}=-34.5$, and $\Delta_{b}^{(1)}/\kappa_{a}=-21$), the three-photon state $|2\rangle_{a}|1\rangle_{b}|-\rangle$ is occupied with a probability 0.024 at time $\kappa_{a}t\approx5.76$, while the other photon state populations are 0 except for the zero-photon state population $P_{|0\rangle_{a}|0\rangle_{b}|+\rangle}(t)$. The dissipation of the cavity field leads to the emission of the first photon, and the wave function collapses into the two-photon state $|1\rangle_{a}|1\rangle_{b}|-\rangle$ with almost unity probability. The second photon and the third photon are further emitted through the cavity field dissipation, and the wave function collapses into the one-photon state $|0\rangle_{a}|1\rangle_{b}|-\rangle$ and the zero-photon state $|0\rangle_{a}|0\rangle_{b}|-\rangle$. This indicates that two-mode correlated three-photon bundle emission can be achieved for the nondegenerate three-photon JC model. Similarly, in Figure \ref{Fig6}c1-c5, d1-d7, we show a part of quantum trajectory for the state populations $P_{|k\rangle_{a}|l\rangle_{b}|\pm\rangle}(t)$ ($k=0, 1$ and $l=0$-$2$) and $P_{|k\rangle_{a}|l\rangle_{b}|\pm\rangle}(t)$ ($k, l=0$-$2$) when $\delta_{a}^{(1)}/\kappa_{a}=16.5$ for the nondegenerate three-photon JC model ($n=1, m=2$) and $\delta_{a}^{(2)}/\kappa_{a}\approx14.48$ for the nondegenerate four-photon JC model ($n=2, m=2$), respectively. It can also be found that the two-mode correlated three-photon (four-photon) bundle emission can be achieved for the nondegenerate three-photon (four-photon) JC model.

\section{Discussions and conclusion}\label{sectionIV}
Finally, we present a brief discussion on the experimental feasibility of the present scheme. In this work, the physical model is a nondegenerate multiphoton JC model, which is composed of a TLS and two bosonic modes. In principle, the physical platform, which could realize the nondegenerate multiphoton JC model, can be used to implement the present scheme. Below, we will analyze how to realize the multiphoton JC model based on both the superconducting quantum circuits\textsuperscript{\cite{Gu2017,Blais2021}} and trapped ion platforms. In the circuit-QED and trapped ion systems, it is possible to realize an interaction with the form $(\hat{\sigma}_{+}+\hat{\sigma}_{-})D_{a}(\alpha)D_{b}(\beta)$ between a TLS and two bosonic modes,\textsuperscript{\cite{Ma2022,Menard2022,Orszag2000}} where $\hat{\sigma}_{+}$ and $\hat{\sigma}_{-}$ are the raising and lowering operators of the TLS, and $D_{a}(\alpha) =e^{\alpha \hat{a}^{\dag }-\alpha ^{\ast }\hat{a}}$ and $D_{b}(\beta) =e^{\beta \hat{b}^{\dag }-\beta ^{\ast }\hat{b}}$ are the displacement operators of the two modes $a$ and $b$, respectively. By expanding the displacement operators, we can obtain the coupling terms as $\hat{\sigma}_{+}\hat{a}^{\dag m}\hat{a}^{m}\hat{b}^{\dag j}\hat{b}^{k}$. Therefore, we can choose proper parameters such that the target terms are near-resonant and other terms are far-off detuned. Then we can obtain the desired interaction Hamiltonian by making the rotating-wave approximation. In addition, we consider the JC-type coupling working in the weak-coupling regime, which is accessible with current experimental techniques.\textsuperscript{\cite{Raimond2001,Blais2021}} Based on these analyses, we know that the present scheme should be within the reach of current experimental conditions.

In conclusion, we have investigated the correlated multiphoton bundle emission in the nondegenerate multiphoton JC model. Specifically, we have focused on the Mollow regime, in which the super-Rabi oscillation between the zero-photon state and the ($n+m$)-photon state can be realized. By investigating the photon number populations, the equal-time high-order correlation function between different cavity modes, the generalized time-delayed second-order correlation functions of photon bundle, and the quantum trajectory of the state populations, we have confirmed the existence of two-mode correlated multiphoton bundle emission. In particular, we have observed the emergence of an antibunching effect among the strongly-correlated photon bundles, confirming the realization of an antibunched ($n+m$)-photon source.


\medskip
\textbf{Appendix: Derivation of the effective oscillating frequency in Eq.~(\ref{Omega}) for the super-Rabi oscillation} \par 

In this Appendix, we present a detailed derivation of the effective oscillating frequency in Equation (\ref{Omega}) for the super-Rabi oscillation. In our system, when the transition $|0\rangle_{a}|0\rangle_{b}|+\rangle \leftrightarrow |n\rangle_{a}|m\rangle_{b}|-\rangle $ is resonant, then the transitions $|n\rangle_{a}|m\rangle_{b}|-\rangle \leftrightarrow |2n\rangle_{a}|2m\rangle_{b}|+\rangle$ and $|n\rangle_{a}|m\rangle_{b}|-\rangle \leftrightarrow |2n\rangle_{a}|2m\rangle_{b}|-\rangle$ in the high-order photon domain will be suppressed by the detunings $2\Omega $ and $\Omega $, respectively. Therefore, the system can be approximately restricted into the subspaces with the bases $\{|0\rangle_{a}|0\rangle_{b}|-\rangle, |0\rangle_{a}|0\rangle_{b}|+\rangle, |1\rangle_{a}|1\rangle_{b}|-\rangle, |1\rangle_{a}|1\rangle_{b}|+\rangle, ...|k\rangle_{a}|l\rangle_{b}|-\rangle, |k\rangle_{a}|l\rangle_{b}|+\rangle,... |n\rangle_{a}|m\rangle_{b}|-\rangle,$\\ $|n\rangle_{a}|m\rangle_{b}|+\rangle\}$. In the absence of dissipation, the wave function of the system at time $t$ can be written as
\begin{equation}
	|\psi(t)\rangle=\sum_{k=0}^{n}\sum_{l=0}^{m}\sum_{s=\pm}d_{k,l,s}(t)|k\rangle_{a}|l\rangle_{b}|s\rangle,  \label{A1}
\end{equation}
where $|k\rangle_{a}|l\rangle_{b}|s\rangle$ represents that there are $k$ ($l$) photons in the optical mode $a$ ($b$) and the TLS is in state $|s\rangle$ for $s=\pm$, with $d_{k,l,s}(t)$ being the corresponding probability amplitude. Based on the Schr\"{o}dinger equation $i|\dot{\psi}(t)\rangle=\hat{H_{I}}|\psi(t)\rangle$ [$\hat{H_{I}}$ is given by Equation (\ref{HI})], the equations of motion for these probability amplitudes can be obtained as
\begin{subequations}\label{A2}
	\begin{align}
		i\dot{d}_{0,0,-}(t) =&E_{-}d_{0,0,-}(t)-g\sqrt{n!m!}c_{-}c_{+}d_{n,m,-}(t)  \notag \\
		&+g\sqrt{n!m!}c_{-}^{2}d_{n,m,+}(t), \label{A2a} \\
		i\dot{d}_{0,0,+}(t) =&E_{+}d_{0,0,+}(t) -g\sqrt{n!m!}c_{+}^{2}d_{n,m,-}(t)   \nonumber \\
		&+g\sqrt{n!m!}c_{+}c_{-}d_{n,m,+}(t), \label{A2b}  \\
		\cdots \nonumber \\
		i\dot{d}_{k,l,-}(t) =&(E_{-}+k\delta_{a}^{(n)}+l\delta_{b}^{(m)})d_{k,l,-}(t) ,\label{A2c}\\
		i\dot{d}_{k,l,+}(t) =&( E_{+}+k\delta _{a}^{(n) }+l\delta_{b}^{(m)}) d_{k,l,+}(t), \label{A2d}\\
		\cdots \nonumber \\
		i\dot{d}_{n,m,-}(t) =&( E_{-}+n\delta _{a}^{(n) }+m\delta_{b}^{(m)}) d_{n,m,-}(t)  -g\sqrt{n!m!}c_{+}^{2}d_{0,0,+}(t)  \nonumber \\
		&-g\sqrt{n!m!}c_{+}c_{-}d_{0,0,-}(t) , \label{A2e}  \\
		i\dot{d}_{n,m,+}(t) =&( E_{+}+n\delta_{a}^{(n)}+m\delta_{b}^{(m)})d_{n,m,+}(t) +g
		\sqrt{n!m!}c_{-}^{2}d_{0,0,-}(t)   \nonumber \\
		&+g\sqrt{n!m!}c_{+}c_{-}d_{0,0,+}(t)  , \label{A2f}
	\end{align}
\end{subequations}
where $k=1, 2,..., n-1$ and $l=1, 2,..., m-1$. 

It can be seen from Equations (\ref{A2}) that the probability amplitudes in the zero- and ($n+m$)-photon domains are coupled with each other, and that the middle equations of motion for $d_{k,l,\pm}(t)$ are decoupled from the other equations. It means that the probability amplitudes $d_{k,l,\pm}(t)$ do not affect the evolution of the states $|0\rangle_{a}|0\rangle_{b}|\pm\rangle$ and $|n\rangle_{a}|m\rangle_{b}|\pm\rangle$. For studying the super-Rabi oscillation between the two states $|0\rangle_{a}|0\rangle_{b}|+\rangle$ and $|n\rangle_{a}|m\rangle_{b}|-\rangle$, we need to adiabatically eliminate the two states $|0\rangle_{a}|0\rangle_{b}|-\rangle$ and $|n\rangle_{a}|m\rangle_{b}|+\rangle$. To this end, we set $\dot{d}_{0,0,-}(t)=\dot{d}_{n,m,+}(t)=0$ in Equations (\ref{A2a}) and (\ref{A2f}). Then $d_{0,0,-}$ and $d_{n,m,+}$ can be obtained as
\begin{subequations}\label{A3}
	\begin{align}
		d_{0,0,-}(t)=&-\frac{g\sqrt{n!m!}(n\delta_{a}^{(n)}+m\delta_{b}^{(m)}+E_{+}) c_{-}c_{+}}{L}d_{n,m,-}(t)   \nonumber \\
		&-\frac{g^{2}n!m!c_{-}^{3}c_{+}}{L}d_{0,0,+}(t), \label{A3a}\\
		d_{n,m,+}(t)=&\frac{g^{2}n!m!c_{-}^{3}c_{+}}{L}d_{n,m,-}(t)+\frac{g\sqrt{n!m!}c_{+}c_{-}E_{-}}{L}d_{0,0,+}(t), \label{A3b}
	\end{align}
\end{subequations}
where we introduce $L=g^{2}n!m!c_{-}^{4}-(n\delta_{a}^{(n)}+m\delta_{b}^{(m)}+E_{+})E_{-}$. By substituting Equations (\ref{A3a}) and ~(\ref{A3b}) into Equations (\ref{A2b}) and (\ref{A2e}), respectively, we obtain the equations of motion for the probability amplitudes $d_{0,0,+}$ and $d_{n,m,-}$ as 
\begin{subequations}\label{A4}
	\begin{align}
		i\dot{d}_{0,0,+}(t)  =&\frac{\sqrt{n!m!}gc_{+}^{2}(n\delta_{a}^{(n)}+m\delta_{b}^{(m)}+E_{+})E_{-}}{n!m!g^{2}c_{-}^{4}-(n\delta_{a}^{(n)}+m\delta_{b}^{(m)}+E_{+})E_{-}} d_{n,m,-}(t) \notag \\ &+\varepsilon _{1}d_{0,0,+}\left( t\right), \\
		i\dot{d}_{n,m,-}(t)=&\frac{\sqrt{n!m!}gc_{+}^{2}(n\delta_{a}^{(n)}+m\delta_{b}^{(m)}+E_{+})E_{-}}{n!m!g^{2}c_{-}^{4}-(n\delta_{a}^{(n)}+m\delta_{b}^{(m)}+E_{+})E_{-}}d_{0,0,+}(t)  \notag \\ &+\varepsilon _{2}d_{n,m,-}(t) ,
	\end{align}
\end{subequations}
where we introduce
\begin{subequations}\label{A6}
	\begin{align}
		\varepsilon_{1}=&\frac{g^{2}n!m!c_{+}^{2}c_{-}^{2}E_{-}}{L}+E_{+},\\
		\varepsilon_{2}=&\frac{g^{2}n!m!c_{+}^{2}c_{-}^{2}(n\delta_{a}^{(n)}+m\delta_{b}^{(m) }+E_{+})}{L}+(n\delta_{a}^{(n)}+m\delta_{b}^{(m)}+E_{-}).
	\end{align}
\end{subequations}
Based on these equations, we can derive an effective Hamiltonian for describing the evolution of the super-Rabi oscillation as
\begin{equation}
	\hat{H}_{\text{eff}}^{\left( n,m\right) }=
	\begin{pmatrix}
		\varepsilon_{1} & \Omega _{\text{eff}}^{(n,m)} \\
		\Omega _{\text{eff}}^{(n,m)} & \varepsilon_{2}
	\end{pmatrix}.\label{Heff}
\end{equation}
Here, we define the bases as $|0\rangle_{a}|0\rangle_{b}|+\rangle=(1,0)^{T}$ and $|n\rangle_{a}|m\rangle_{b}|-\rangle =(0,1)^{T}$ (the superscript $\text{\textquotedblleft }T\text{\textquotedblright}$ denoting the matrix transpose). The effective oscillating frequency $\Omega_{\text{eff}}^{(n,m)}$ of the super-Rabi oscillation is given by Equation (\ref{Omega}).

\medskip
\textbf{Acknowledgements} \par 
J.-Q.L. was supported in part by National Natural Science Foundation of China (Grants No. 12175061, No. 12247105, and No. 11935006), the Science and Technology Innovation Program of Hunan Province (Grants No. 2021RC4029 and No. 2020RC4047). F.Z. was supported in part by the China Postdoctoral Science Foundation (Grant No. 2021M700360).


\begin{thebibliography}{73}%
\providecommand{\url}[1]{\texttt{#1}}
\providecommand{\urlprefix}{URL }

\bibitem{Dell20o6}      F. Dell'Anno, S. De Siena, F. Illuminati, \emph{Phys. Rep.} \textbf{2006}, \emph{428}, 53.

\bibitem{Pan2012}      J.-W. Pan, Z.-B. Chen, C.-Y. Lu, H. Weinfurter, A. Zeilinger, M. Zukowski, \emph{Rev. Mod. Phys.} \textbf{2012}, \emph{84}, 777.

\bibitem{Giovannetti2004}     V. Giovannetti, S. Lloyd, L. Maccone, \emph{Science} \textbf{2004}, \emph{306}, 1330.

\bibitem{Giovannetti2006}     V. Giovannetti, S. Lloyd, L. Maccone, \emph{Phys. Rev. Lett.} \textbf{2006}, \emph{96}, 010401.

\bibitem{PhysRevLett2000}     A. N. Boto, P. Kok, D. S. Abrams, S. L. Braunstein, C. P. Williams, J. P. Dowling, \emph{Phys. Rev. Lett.} \textbf{2000}, \emph{85}, 2733.

\bibitem{Two-photon2001}     M. D$^{\prime }$Angelo, M. V. Chekhova, Y. Shih, \emph{Phys. Rev. Lett.} \textbf{2001}, \emph{87}, 013602.

\bibitem{López2015}      J. C. L. Carre\~{n}o, C. S. Mu\~{n}oz, D. Sanvitto, E. del Valle, F. P. Laussy, \emph{Phys. Rev. Lett.} \textbf{2015}, \emph{115}, 196402.

\bibitem{Dorfman2016}      K. E. Dorfman, F. Schlawin, S. Mukamel, \emph{Rev. Mod. Phys.} \textbf{2016}, 88, 045008.

\bibitem{Moreau2019}      P.-A. Moreau, E. Toninelli, T.
Gregory, M. J. Padgett, \emph{Nat. Rev. Phys.} \textbf{2019}, \emph{1}, 367.

\bibitem{Kimble2008Jun}     H. J. Kimble, \emph{Nature} \textbf{2008}, \emph{453}, 1023.

\bibitem{Gisin2006}      N. Gisin, R. Thew, \emph{Nature Photonics,} \textbf{2006}, \emph{1}, 165.

\bibitem{Gisin2002}      N. Gisin, G. Ribordy, W. Tittel, H. Zbinden, \emph{Rev. Mod. Phys.} \textbf{2002}, \emph{74}, 145.

\bibitem{alez-Tudela2017}      A. Gonz\'{a}ez-Tudela, V. Paulisch, H. J. Kimble, J. I. Cirac, \emph{Phys. Rev. Lett.} \textbf{2017}, \emph{118}, 213601.

\bibitem{Bienias2014}      P. Bienias, S. Choi, O. Firstenberg, M. F. Maghrebi, M. Gullans, M. D. Lukin, A. V. Gorshkov, H. P. B\"{u}chler, \emph{Phys. Rev. A} \textbf{2014}, \emph{90}, 053804.

\bibitem{Maghrebi2015}      M. F. Maghrebi, M. J. Gullans, P. Bienias, S. Choi, I. Martin, O. Firstenberg, M. D. Lukin, H. P. B\"{u}chler, A. V. Gorshkov, \emph{Phys. Rev. Lett.} \textbf{2015}, \emph{115}, 123601.

\bibitem{Law1996}      C. K. Law, J. H. Eberly, \emph{Phys. Rev. Lett.} \textbf{1996}, \emph{76}, 1055.

\bibitem{Huang2014}      J.-F. Huang, C. K. Law, \emph{Phys. Rev. A} \textbf{2014}, \emph{89}, 033827.

\bibitem{Dousse2010}       A. Dousse, J. Suffczy\'{n}ski, A. Beveratos, O. Krebs, A. Lema\^{i}tre, I. Sagnes, J. Bloch, P. Voisin, P. Senellart, \emph{Nature} \textbf{2010}, \emph{466}, 217.

\bibitem{Iwamoto2011}       Y. Ota, S. Iwamoto, N. Kumagai, Y. Arakawa, \emph{Phys. Rev. Lett}. \textbf{2011}, \emph{107}, 233602.

\bibitem{Koshino2013}        K. Koshino, K. Inomata, T. Yamamoto, Y. Nakamura, \emph{Phys. Rev. Lett}. \textbf{2013}, \emph{111}, 153601.

\bibitem{Photonics2014}      M. M\"{u}ller, S. Bounouar, K. D. J\"{o}ns, M. Gl\"{a}ssl, P. Michler, \emph{Nat. Photonics} \textbf{2014}, \emph{8}, 224.

\bibitem{Chang2016}      Y. Chang, A. Gonz\'{a}lez-Tudela, C. S. Mu\~{n}oz, C. Navarrete-Benlloch, T. Shi, \emph{Phys. Rev. Lett.} \textbf{2016}, \emph{117}, 203602.

\bibitem{Hargart2016}      F. Hargart, M. M\"{u}ller, K. Roy-Choudhury, S. L. Portalupi, C. Schneider, S. H\"{o}fling, M. Kamp, S. Hughes, P. Michler, \emph{Phys. Rev. B} \textbf{2016}, \emph{93}, 115308.

\bibitem{Zueco2016}       E. S\'{a}nchez-Burillo, L. Mart\'{i}n-Moreno, J. J. Garc\'{i}a-Ripoll, D. Zueco, \emph{Phys. Rev. A} \textbf{2016}, \emph{94}, 053814.

\bibitem{Liao2010}       J.-Q. Liao, C. K. Law, \emph{Phys. Rev. A} \textbf{2010}, \emph{82}, 053836.

\bibitem{Liao2013}       J.-Q. Liao, C. K. Law, \emph{Phys. Rev. A} \textbf{2013}, \emph{87}, 043809.

\bibitem{Qin2019}       W. Qin, V. Macr\`{i}, A. Miranowicz, S. Savasta, F. Nori, \emph{Phys. Rev. A} \textbf{2019}, \emph{100}, 062501.

\bibitem{Laussy2014}     C. S. Mu\~{n}oz, E. del Valle, A. G. Tudela, K. M\"{u}ller, S. Lichtmannecker, M. Kaniber, C. Tejedor, J. J. Finley, F. P. Laussy, \emph{Nat. Photonics} \textbf{2014}, \emph{8}, 550.

\bibitem{SanchezMunoz2018}     C. S. Mu\~{n}oz, F. P. Laussy, E. del Valle, C. Tejedor, A. Gonz\'{a}lez-Tudela, \emph{Optica} \textbf{2018}, \emph{5}, 14.

\bibitem{Bin2021}     Q. Bin, Y. Wu, X.-Y. L\"{u}, \emph{Phys. Rev. Lett.} \textbf{2021}, \emph{127}, 073602.


\bibitem{liu2022}     C. Liu, J.-F. Huang, L. Tian, \emph{Sci. China: Phys., Mech. Astron.} \textbf{2022}, \emph{66}, 220311.

\bibitem{Cosacchi2021}     M. Cosacchi, A. Mielnik-Pyszczorski, T. Seidelmann, M. Cygorek, A. Vagov, D. E. Reiter, V. M. Axt, \emph{Phys. Rev. B} \textbf{2022}, \emph{106}, 115304.

\bibitem{Gou2022}     C. Gou, X. Hu, F. Wang, \emph{Phys. Rev. A} \textbf{2022}, \emph{106}, 063718.

\bibitem{Ren2022}     Y. Ren, Z. L. Duan, B. X. Fan, S. G. Guan, M. Xie, C. J. Liu, \emph{Opt. Express} \textbf{2022}, \emph{30}, 21787.

\bibitem{Jiang2022}     S. Y. Jiang, F. Zou, Y. Wang, J. F. Huang, X. W. Xu, J.-Q. Liao, \emph{Opt. Express} \textbf{2023}, \emph{31}, 15697.

\bibitem{Ma2021}     S.-L. Ma, X.-K. Li, Y.-L. Ren, J.-K. Xie, F.-L. Li, \emph{Phys. Rev. Res.} \textbf{2021}, \emph{3}, 043020.

\bibitem{Ma2022}     S.-L. Ma, J.-K. Xie, Y.-L. Ren, X.-K. Li, F.-L. Li, \emph{New J. Phys.} \textbf{2022}, \emph{24}, 053001.

\bibitem{Zou2023}     F. Zou, Y. Li, J.-Q. Liao, \emph{New J. Phys.} \textbf{2023}, \emph{25}, 043027.

\bibitem{Bin2020}     Q. Bin, X.-Y. L\"{u}, F. P. Laussy, F. Nori, Y. Wu, \emph{Phys. Rev. Lett.} \textbf{2020}, \emph{124}, 053601.

\bibitem{Deng2021}     Y. G. Deng, T. Shi, S. Yi, \emph{Photonics
Res.} \textbf{2021}, \emph{9}, 1289.

\bibitem{Zou2022}     F. Zou, J.-Q. Liao, Y. Li, \emph{Phys. Rev. A} \textbf{2022,} \emph{105}, 053507.

\bibitem{Menard2022}     G. C. M\'{e}nard, A. Peugeot, C. Padurariu, C. Rolland, B. Kubala, Y. Mukharsky, Z. Iftikhar, C. Altimiras, P. Roche, H. le Sueur, P. Joyez, D. Vion, D. Esteve, J. Ankerhold, F. Portier, \emph{Phys. Rev. X} \textbf{2022}, \emph{12}, 021006.

\bibitem{Ridolfo2012}       A. Ridolfo, M. Leib, S. Savasta, M. J. Hartmann, \emph{Phys. Rev. Lett.} \textbf{2012}, \emph{109}, 193602.

\bibitem{Miranowicz2013}       A. Miranowicz, M. Paprzycka, Y.-X. Liu, J. Bajer, F. Nori, \emph{Phys. Rev. A} \textbf{2013}, \emph{87}, 023809.

\bibitem{Huang2018}       R. Huang, A. Miranowicz, J.-Q. Liao, F. Nori, H. Jing, \emph{Phys. Rev. Lett.} \textbf{2018}, \emph{121}, 153601.

\bibitem{Zou2020}       F. Zou, X.-Y. Zhang, X.-W. Xu, J.-F. Huang, J.-Q. Liao, \emph{Phys. Rev. A} \textbf{2020}, \emph{102}, 053710.

\bibitem{Ren2021}       Y. Ren, S. H. Duan, W. Z. Xie, Y. K. Shao, Z. L. Duan, \emph{Phys. Rev. A} \textbf{2021}, \emph{103}, 053710.

\bibitem{Ashraf1992}       M. M. Ashraf, M. S. K. Razmi, \emph{Phys. Rev. A} \textbf{1992}, \emph{45}, 8121.

\bibitem{Ashraf1994}       M. M. Ashraf, \emph{Phys. Rev. A} \textbf{1994}, \emph{50}, 5116.

\bibitem{Mollow1969}     B. R. Mollow, \emph{Phys. Rev.} \textbf{1969}, \emph{188}, 1969.


\bibitem{Kimble1976}     H. J. Kimble, L. Mandel, \emph{Phys. Rev. A} \textbf{1976}, \emph{13}, 2123.

\bibitem{Ulhaq2012}     A. Ulhaq, S. Weiler, S. M. Ulrich, R. Ro\ss bach, M. Jetter, P. Michler, \emph{Nat. Photonics} \textbf{2012}, \emph{6}, 238.

\bibitem{Gonzalez-Tudela2013}     A. Gonz\'{a}lez-Tudela, F. P. Laussy, C. Tejedor, M. J. Hartmann, E. del Valle, \emph{New J. Phys.} \textbf{2013}, \emph{15}, 033036.

\bibitem{Carreño2017}     J. C. L. Carre\~{n}o, E. del Valle, F. P. Laussy, \emph{Laser Photonics Rev.} \textbf{2017}, \emph{11}, 1700090.

\bibitem{Johansson2012}     J. Johansson, P. Nation, F. Nori, \emph{Comput. Phys. Commun}, \textbf{2012}, \emph{183}, 1760.

\bibitem{Johansson2013}     J. Johansson, P. Nation, F. Nori, \emph{Comput. Phys. Commun} \textbf{2013}, \emph{184}, 1234.

\bibitem{Stassi2013}     R. Stassi, A. Ridolfo, O. Di Stefano, M. J. Hartmann, S. Savasta, \emph{Phys. Rev. Lett.} \textbf{2013}, \emph{110}, 243601.

\bibitem{Ou2017}     Z. J. Ou, \emph{Quantum Optics for Experimentalists}, World Scientific, Singapore, \textbf{2017}.

\bibitem{Mandel1995}     L. Mandel, E. Wolf, \emph{Optical Coherence and Quantum Optics}, Cambridge University Press, Cambridge,\textbf{1995}.

\bibitem{Gu2017}     X. Gu, A. F. Kockum, A. Miranowicz, Y. X. Liu, F. Nori, \emph{Physics Reports} \textbf{2017}, \emph{718-719}, 1.

\bibitem{Blais2021}     A. Blais, A. L. Grimsmo, S. M. Girvin, A. Wallraff, \emph{Rev. Mod. Phys.} \textbf{2021}, \emph{93}, 025005.

\bibitem{Orszag2000}     M. Orszag, \emph{Quantum Optics: Including Noise Reduction, Trapped Ions, Quantum Trajectories, and Decoherence}, Springer-Verlag, Switzerland, \textbf{2000}.

\bibitem{Raimond2001}     J. M. Raimond, M. Brune, S. Haroche, \emph{Rev. Mod. Phys.} \textbf{2001}, \emph{73}, 565.






\end{thebibliography}
\end{document}